\documentclass[12pt,floatfix,nofootinbib,superscriptaddress,aps,prd]{revtex4-2}
\usepackage[utf8]{inputenc}
\usepackage{float}
\usepackage{array}
\usepackage{lipsum}
\usepackage{graphicx}
\usepackage{amsmath,amsthm,amsfonts,amssymb}
\usepackage{tikz}
\usepackage{multirow}
\usepackage{braket}
\newcommand{\be}{\begin{equation}}
\newcommand{\ee}{\end{equation}}
\newcommand{\Be}{\begin{eqnarray}}
\newcommand{\Ee}{\end{eqnarray}}

\newcommand{\mincir}{\raise
-3.truept\hbox{\rlap{\hbox{$\sim$}}\raise4.truept\hbox{$<$}\ }}
\newcommand{\magcir}{\raise
-3.truept\hbox{\rlap{\hbox{$\sim$}}\raise4.truept\hbox{$>$}\ }}

\newcolumntype{Y}{>{\centering\arraybackslash}X}
\providecommand{\U}[1]

 \usetikzlibrary{quotes,angles}
 \usetikzlibrary{arrows} 
\usetikzlibrary{decorations.markings}
\usepackage{graphicx}
\usepackage[english]{babel}
\usepackage{color}
\usepackage{subfig}
\usepackage{caption}
\usepackage{tensor}
\usepackage{esint}
\usepackage[dvips]{epsfig}
\usepackage[dvips]{graphicx}
\usepackage{float}
\usepackage{units}
\usepackage{textcomp}
\usepackage{mathrsfs}
\usepackage{amsmath}
\usepackage[makeroom]{cancel}
\usepackage{amssymb}
\usepackage{amsbsy}
\usepackage{amsfonts}
\usepackage{amssymb,mathrsfs,xcolor}
\usepackage{esint}
\usepackage{array}
\usepackage{graphicx}

\usepackage{hyperref}
\hypersetup{
    colorlinks,
    citecolor=blue,
    filecolor=green,
    linkcolor=purple,
    urlcolor=red,
}

\usepackage{slashed}

\newcommand{\ie}{\begin{equation}}
\newcommand{\fe}{\end{equation}}
\newcommand{\se}{\begin{eqnarray}}
\newcommand{\ff}{\end{eqnarray}}

\usepackage{hyperref}
\hypersetup{colorlinks,breaklinks,
			citecolor=[rgb]{0,0.0,1.0},
            urlcolor=[rgb]{0.0,0.0,1.0},
            linkcolor=[rgb]{0,0.5,0.9}}

\begin{document}

\title{Geodesics, accretion disk, gravitational lensing, time delay, and effects on neutrinos induced by a non--commutative black hole}


\author{A. A. Ara\'{u}jo Filho}
\email{dilto@fisica.ufc.br}
\affiliation{Departamento de Física, Universidade Federal da Paraíba, Caixa Postal 5008, 58051--970, João Pessoa, Paraíba,  Brazil.}


\author{N. Heidari}
\email{heidari.n@gmail.com}

\affiliation{Center for Theoretical Physics, Khazar University, 41 Mehseti str.,  Baku, AZ 1096, Azerbaijan.}
\affiliation{School of Physics, Damghan University,
Damghan, 3671641167, Iran} 

\author{Ali \"Ovg\"un}
\email{ali.ovgun@emu.edu.tr}
\affiliation{Physics Department, Eastern Mediterranean University, Famagusta, 99628 North
Cyprus via Mersin 10, Turkiye.}


\date{\today}

\begin{abstract}

This paper explores gravitational phenomena associated with a non--commutative black hole. Geodesic equations are derived, and a thin accretion disk is analyzed to model the black hole shadow image, considering an optically thin, radiating, and infalling gas. Retrolensing effects are examined to trace photon emission configurations, while gravitational lensing is investigated through weak and strong deflection limits, with lensing equations and observables applied to Sagittarius A*. The study also includes calculations of time delay, energy deposition rate from neutrino annihilation, phase and probability of neutrino oscillation, and neutrino gravitational lensing.

\end{abstract}


\maketitle

\pagebreak

\tableofcontents

\pagebreak

\section{Introduction}

Gravity, as described by general relativity, is expressed through a geometric framework that inherently involves nonlinear dynamics. This nonlinearity presents substantial challenges in obtaining exact solutions to the Einstein field equations, even under the imposition of specific symmetries or constraints \cite{wald2010general, misner1973gravitation}. To manage these challenges, the weak-field approximation is widely adopted. This method simplifies the equations, allowing for the study of gravitational waves, which emerge as a significant feature of the theory. These waves are instrumental in understanding black hole phenomena, including their stability, emission of \textit{Hawking} radiation, and interactions with their astrophysical surroundings.

The framework of general relativity, which describes the geometry of spacetime, does not inherently impose limits on the precision of distance measurements. However, it is widely hypothesized that such precision is fundamentally constrained by the Planck length. To address this theoretical challenge, non-commutative spacetime models have been introduced. Originating from developments in string theory \cite{3,szabo2003quantum,szabo2006symmetry}, these models have also found significant applications in supersymmetric Yang--Mills theories \cite{ferrari2004superfield,ferrari2003finiteness,ferrari2004towards}. In gravitational contexts, non-commutativity is frequently incorporated through the Seiberg--Witten map by gauging appropriate symmetry groups \cite{chamseddine2001deforming}.

The application of non--commutative geometry has resulted in substantial advancements in black hole research \cite{Anacleto:2019tdj,anacleto2023absorption,anacleto2021quasinormal,heidari2024exploring,1,2,modesto2010charged,mann2011cosmological,nicolini2009noncommutative,zhao2023quasinormal,karimabadi2020non,campos2022quasinormal,lopez2006towards}, including investigations of black hole evaporation \cite{myung2007thermodynamics,23araujo2023thermodynamics} and their thermodynamic properties \cite{sharif2011thermodynamics,banerjee2008noncommutative,nozari2007thermodynamics,lopez2006towards,nozari2006reissner}. Furthermore, the thermal properties of field theories within non--commutative frameworks have been examined in diverse scenarios \cite{furtado2023thermal,araujo2023thermodynamical}.

The concept of non--commutative spacetime, a fundamental idea in modern theoretical physics, is represented by the commutation relation \( [x^\mu, x^\nu] = i \Theta^{\mu \nu} \). Here, \( x^\mu \) are spacetime coordinates, and \( \Theta^{\mu \nu} \) is an anti--symmetric tensor that encapsulates the non--commutative structure. Integrating non--commutativity into gravitational theories has been a focus of various methodologies. One prominent strategy employs the non--commutative gauge group SO(4,1), associated with de Sitter (dS) symmetry, in combination with the Poincaré group ISO(3,1), facilitated by the Seiberg–Witten (SW) map. Chaichian et al. \cite{chaichian2008corrections} utilized this framework to construct a deformed Schwarzschild metric, incorporating the effects of spacetime non--commutativity.

An innovative approach introduced by Nicolini et al. \cite{nicolini2006noncommutative} integrates the effects of spacetime non-commutativity into general relativity by modifying the matter source term in the Einstein field equations. Instead of altering the Einstein tensor, this method replaces the conventional point-like mass density with a non-singular distribution. Two specific forms are employed: a Gaussian profile, \( \rho_\Theta = M (4\pi \Theta)^{-\frac{3}{2}} e^{-\frac{r^2}{4\Theta}} \), and a Lorentzian distribution, \( \rho_\Theta = M \sqrt{\Theta} \pi^{-\frac{3}{2}} (r^2 + \pi \Theta)^{-2} \).

The detection of gravitational waves by experiments such as LIGO and Virgo \cite{016,017,018} has opened new frontiers in cosmological research. These waves now serve as crucial probes for investigating the universe, including gravitational lensing phenomena within the framework of the weak-field approximation \cite{019,020}. Historically, studies on gravitational lensing have concentrated on the deflection of light over cosmic distances, often modeled in Schwarzschild spacetime \cite{021}. This work was later generalized to include spherically symmetric and static spacetimes \cite{022}. However, in regions dominated by intense gravitational fields, such as the vicinity of black holes, the angular deflection of light becomes notably more pronounced, a behavior consistent with expectations under strong-field conditions.

The groundbreaking imaging of the supermassive black hole at the center of the M87 galaxy by the Event Horizon Telescope has drawn substantial scientific attention \cite{023,024,025,026,027,028,029}. Earlier, Virbhadra and Ellis developed a simplified lens equation for analyzing the gravitational lensing effects of supermassive black holes within asymptotically flat spacetimes \cite{030,031}. Their findings revealed that strong gravitational fields near such massive objects produce multiple symmetrically distributed images along the optical axis.

This foundational work was later extended through analytical advancements by Fritelli et al. \cite{032}, Bozza et al. \cite{033}, and Tsukamoto \cite{035}, who refined the methods used to investigate gravitational lensing in the strong-field regime. These studies examined the bending of light in various scenarios, including Schwarzschild spacetime \cite{virbhadra2002gravitational,virbhadra1998role,virbhadra2000schwarzschild,grespan2023strong,cunha2018shadows,oguri2019strong,metcalf2019strong,bisnovatyi2017gravitational,ezquiaga2021phase,Okyay:2021nnh,Ovgun:2018fnk,Ovgun:2018tua,Li:2020dln,Pantig:2022gih,Pantig:2022ely,Kuang:2022xjp,Donmez:2024lfi,Donmez:2023wtf,Donmez:2023egk,Koyuncu:2014nga}, as well as in exotic structures like wormholes \cite{ovgun2019exact,38.1,38.2,38.3,38.4,38.5}, rotating black hole solutions \cite{hsieh2021strong,hsieh2021gravitational,jusufi2018gravitational,37.1,37.2,37.3,37.4,37.5,37.6}, and frameworks grounded in alternative gravitational theories \cite{chakraborty2017strong,40,nascimento2024gravitational}. Lens effects in Reissner--Nordström spacetimes \cite{036,036.1,036.2} and other configurations \cite{zhang2024strong,tsukamoto2023gravitational} have also been explored. Additionally, studies addressing gravitational distortions have provided further information about the impact of extreme gravitational fields on light propagation \cite{virbhadra2024conservation,virbhadra2022distortions}.

This study focuses on gravitational phenomena linked to a non--commutative black hole. The geodesic equations are formulated, and a thin accretion disk model is employed to simulate the black hole's shadow image, incorporating an optically thin, radiating, and infalling gas. Retrolensing is analyzed to map photon emission configurations, while gravitational lensing is explored in both weak and strong deflection regimes, with the lensing equations and observables applied specifically to Sagittarius A*. Additionally, the analysis includes calculations of the time delay, the energy deposition rate from neutrino annihilation, the phase and probability of neutrino oscillations, as well as the impact of gravitational lensing on neutrinos.


\section{The black hole solution}

The study presented in Ref. \cite{9} proposed a framework for describing gravitational field deformation by utilizing the non--commutative de Sitter group, SO(4,1), and implementing the Seiberg--Witten map. The contraction of the SO(4,1) group to the Poincaré group, ISO(3,1), enabled the derivation of modified gravitational gauge potentials, commonly referred to as tetrad fields, denoted by \( \hat{e}_\mu^a(x, \Theta) \). These deformed tetrad fields were then applied to the Schwarzschild spacetime, resulting in the development of a modified Schwarzschild metric that incorporates the effects of non--commutativity up to second order. The metric is expressed as:
\begin{equation}\label{met}
\begin{array}{l}
{{g}_{tt}^{\Theta}} = {g_{tt}} - \frac{{\alpha (8r - 11\alpha )}}{{16{r^4}}}{\Theta ^2} + ...,\\
{{g}^{\Theta}_{rr}} = {g_{rr}} - \frac{{\alpha (4r - 3\alpha )}}{{16{r^2}{{(r - \alpha )}^2}}}{\Theta ^2} + ...,\\
{{g}^{\Theta}_{\theta\theta}} = {g_{\theta\theta}} - \frac{{2{r^2} - 17\alpha r + 17{\alpha ^2}}}{{32r(r - \alpha )}}{\Theta ^2} + ...,\\
{{g}^{\Theta}_{\varphi\varphi }} = {g_{\varphi\varphi}} - \frac{{({r^2} + \alpha r - {\alpha ^2})\cos \theta  - \alpha (2r - \alpha )}}{{16r(r - \alpha )}}{\Theta ^2} + ... \, .
\end{array}
\end{equation}
This work defines the parameter $\alpha$ as $\alpha = 2M$, $M$ corresponding to the black hole mass. The deformed metric tensor, denoted by $g^{\Theta}_{\mu \nu}$, is formulated in spherical coordinates, with $g_{\mu \nu}$ serving as the standard Schwarzschild metric. To calculate the radius of the deformed Schwarzschild event horizon, $1/g^{\Theta}_{rr} = 0$ is imposed, i.e., up to the second order of $\Theta$. This yields the modified event horizon radius
\ie
\label{rad}
{r_{s}^{\Theta}} = 2M - \frac{{{\Theta ^2}}}{{32M}}.
\fe
The deformed Schwarzschild black hole features a radius, \( r_{s\Theta} = 2M_\Theta \), which corresponds to the modified non--commutative (NC) mass. This introduces a redefined mass parameter expressed as  
\cite{heidari2023gravitational,araujo2024quantum}:
\begin{equation}\label{mass}
 M_{\Theta} = M - \frac{1}{64M}\Theta^2.
\end{equation}
This study utilizes the standard Schwarzschild metric in conjunction with the deformed non--commutative mass parameter outlined in Eq. (\ref{mass}).


\section{Geodesics}

Geodesics play a crucial role in physics by describing spacetime curvature and the trajectories of particles under gravitational influence. In NC scenarios, their study has become an important area of research, focusing on the quantum effects that alter spacetime properties. Additionally, analyzing the geodesic structure of NC black holes is key to interpreting astrophysical phenomena, such as the dynamics of accretion disks and the characteristics of black hole shadows. In this context, this section is dedicated to conducting such an investigation. The geodesic equation is expressed as:
\ie
\frac{\mathrm{d}^{2}x^{\mu}}{\mathrm{d}s^{2}} + \Gamma\indices{^\mu_\alpha_\beta}\frac{\mathrm{d}x^{\alpha}}{\mathrm{d}s}\frac{\mathrm{d}x^{\beta}}{\mathrm{d}s} = 0, \label{geodesicfull}
\fe
In this framework, \( s \) is introduced as an arbitrary parameter. The main focus is to analyze the impact of non--commutativity on the paths of massless particles. This analysis involves solving a complex system of partial differential equations derived from Eq. (\ref{geodesicfull}). The formulation produces four interdependent partial differential equations that must be addressed to fully explore the influence of non--commutative effects on particle trajectories 
\ie
t'' = \frac{\left(\Theta ^2-64 M^2\right) r' t'}{r \left(\Theta ^2-64 M^2+32 M r\right)},
\fe
\ie
\begin{split}
r'' = & \frac{\left(-\Theta ^2+64 M^2-32 M r\right) \left(\left(64 M^2-\Theta ^2\right) \left(t'\right)^2-64 M r^3 \left(\left(\theta '\right)^2+\sin ^2(\theta ) \left(\varphi '\right)^2\right)\right)}{2048 M^2 r^3}\\
& +\frac{\left(64 M^2-\Theta ^2\right) \left(r'\right)^2}{2 r \left(\Theta ^2-64 M^2+32 M r\right)},
\end{split}
\fe
\ie
\theta'' = \sin (\theta ) \cos (\theta ) \left(\varphi'\right)^2-\frac{2 \theta' r'}{r},
\fe
\ie
\varphi'' = -\frac{2 \varphi' \left(r'+r \theta' \cot (\theta )\right)}{r},
\fe

\begin{figure}
    \centering
     \includegraphics[scale=0.45
     ]{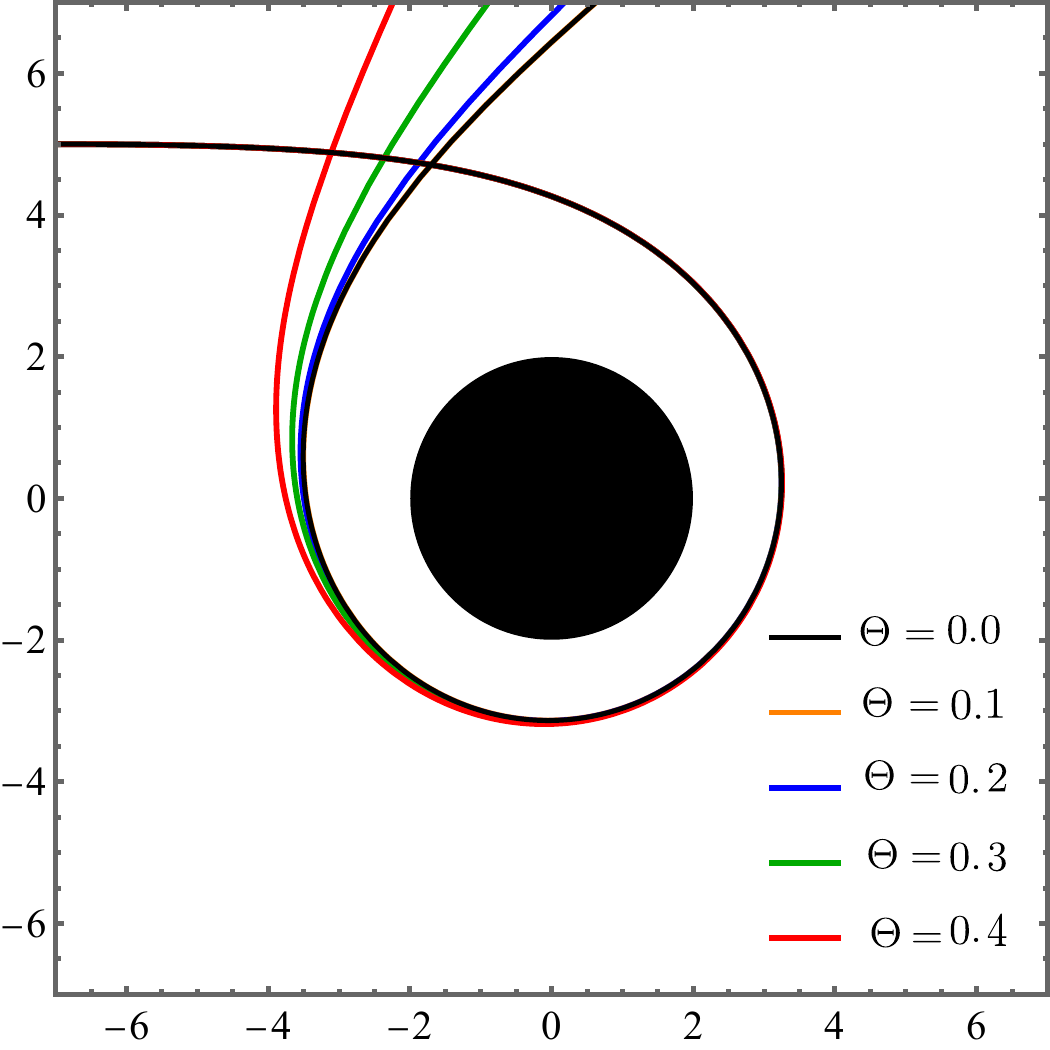}
    \includegraphics[scale=0.45]{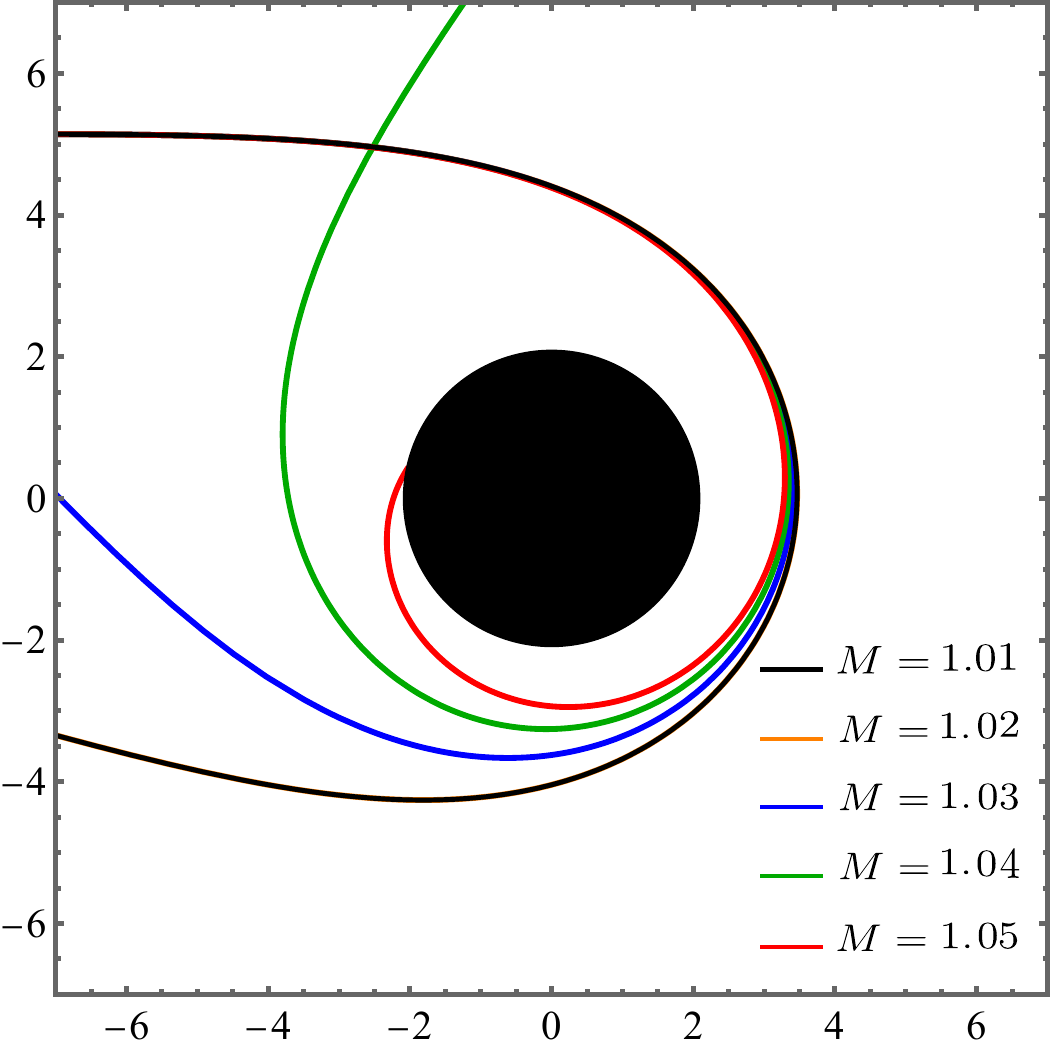}
    \caption{Geodesic trajectories are computed for different values of \(\Theta\) and \(M\).  In the left panel $M = 1$ and in the right panel $\Theta = 0.1$.}
    \label{fig:masslessgeodesics}
\end{figure}

The prime symbol $\prime$ indicates differentiation with respect to \( s \), namely, \(\mathrm{d}/\mathrm{d}s\). Fig. \ref{masslessgeodesics} presents the numerically computed geodesic trajectories for different values of \(\Theta\) and \(M\). The big black disk represents the black hole. It is evident from the Fig. \ref{fig:masslessgeodesics} that the higher non--commutativity parameter lowers the gravitational lensing. However, when $M$ increases for a fixed value of $\Theta$, the gravitational lensing effect gets more powerful.

\section{Thin--Accretion disk}

\begin{table}[ht]
\centering
\caption{Photon Sphere Radii (\(r_{ph}\)) and Shadow Radii (\(r_{sh}\)) vs \(\Theta/M\)}
\begin{tabular}{ccc}
\hline
\(\Theta/M\) & \(r_{ph}\) & \(r_{sh}\) \\
\hline
0.001000 & 3.000000 & 5.196153 \\
0.106211 & 3.000529 & 5.197068 \\
0.211421 & 3.002095 & 5.199782 \\
0.316632 & 3.004699 & 5.204292 \\
0.421842 & 3.008341 & 5.210600 \\
0.527053 & 3.013021 & 5.218706 \\
0.632263 & 3.018739 & 5.228609 \\
0.737474 & 3.025494 & 5.240309 \\
0.842684 & 3.033287 & 5.253807 \\
0.947895 & 3.042117 & 5.269102 \\
1.053105 & 3.051986 & 5.286194 \\
1.158316 & 3.062892 & 5.305085 \\
1.263526 & 3.074836 & 5.325772 \\
1.368737 & 3.087818 & 5.348257 \\
1.473947 & 3.101837 & 5.372539 \\
1.579158 & 3.116894 & 5.398619 \\
1.684368 & 3.132989 & 5.426496 \\
1.789579 & 3.150122 & 5.456171 \\
1.894789 & 3.168292 & 5.487643 \\
2.000000 & 3.187500 & 5.520912 \\
\hline
\end{tabular}
\label{tab:photon_sphere_radii}
\end{table}

In this section, it is investigated an accretion flow around a compact object, characterized by a radiating, optically thin medium. To analyze the resulting shadow produced by such a flow, we adopt a numerical framework grounded in the Backward Raytracing method~\cite{Bambi:2012tg, Okyay:2021nnh}. The derivation of the intensity map for the emitting region involves specific assumptions about the nature of the radiative processes and emission mechanisms. The specific intensity \( I_{\nu_0} \) observed at the photon frequency \( \nu_\text{obs} \), corresponding to the position \((X, Y)\) on the observer's plane, is quantified in units of \(\text{erg}~\text{s}^{-1}~\text{cm}^{-2}~\text{str}^{-1}~\text{Hz}^{-1}\) as follows:
\ie
I_\text{obs}(\nu_\text{obs}, X, Y) = \int_\gamma g^3 j(\nu_e) \, \mathrm{d}l_\text{prop}.
\fe
Here, the redshift factor, denoted as \( g = \nu_\text{obs}/\nu_e \), quantifies the ratio between the photon frequency observed, \( \nu_\text{obs} \), and that measured in the emitter's rest frame, \( \nu_e \). The emissivity per unit volume in the emitter's rest frame is represented by \( j(\nu_e) \), while \( dl_\text{prop} = k_\alpha u^\alpha \, d\lambda \) defines the infinitesimal proper length in the emitter's rest frame. The computation of the redshift factor involves the following relation:
\ie
g = \frac{k_\alpha u^\alpha_\text{obs}}{k_\beta u^\beta_e}.
\fe
In this framework, \( k_\mu \) describes the photon's four--momentum, while \( u^\alpha_e \) corresponds to the four--velocity of the radiating accretion flow. The observer's four--velocity is specified as \( u^\mu_\text{obs} = (1, 0, 0, 0) \), and \( \lambda \) serves as the affine parameter tracing the photon's trajectory, \(\gamma\). The path integral indicated by \(\gamma\) signifies that the computation is performed along the null geodesics traversed by the photon.

The motion of the gas is assumed to follow a radial free--fall trajectory. In the context of a static, spherically symmetric spacetime, this behavior can be described by a specific form of the four-velocity, which simplifies to:
\ie
u^t_e = \frac{1}{g_{tt}(r)}, \quad
    u^r_e = -\sqrt{\frac{1 - g_{tt}(r)}{g_{tt}(r) g_{rr}(r)}},
    \quad u^\theta_e = 0, \quad
    u^\phi_e = 0.
\fe
To address further computations, a connection is now formulated between the time and radial components of the photon's four--velocity:
\ie
    \frac{k_r}{k_t} = \pm \sqrt{g_{rr} \left( \frac{1}{g_{tt}} - \frac{b^{2}}{g_{\phi \phi}} \right)},
\fe
in which $\pm$ sign designates the photon's trajectory, with $+$ corresponding to motion away from the massive object and \(-\) representing motion toward it; and $b$ denotes the impact parameter. Based on this, the redshift factor $g$ can be written as:
\ie
g = \frac{1}{\sqrt{\frac{1}{g_{tt}} \pm \frac{k_r}{k_t} \sqrt{\frac{1 - g_{tt}}{g_{tt} g_{rr}}}}}.
\fe

To describe the specific emissivity, a simplified model is employed. In this approach, the emission is assumed to be monochromatic at a fixed frequency \(\nu_\star\) in the emitter's rest frame and follows a radial distribution that scales as \(1/r^2\):
\ie
j(\nu_e) \propto \frac{\delta(\nu_e - \nu_\star)}{r^2}.
\fe
Here, \(\delta\) represents the Dirac delta function. The proper length is formulated as:
\ie
\mathrm{d} l_\text{prop} = k_\alpha u^\alpha_e \mathrm{d} \lambda = -\frac{k_t}{g |k_r|} \mathrm{d}r.
\fe

By integrating the specific intensity across all observed frequencies, the total observed flux can be determined as:
\ie
F_\text{obs}(X, Y) \propto -\int_\gamma \frac{g^3 k_t}{r^2 k_r} \mathrm{d}r,
\fe
This result will be applied to generate shadow images of the black hole, specifically considering the non-commutative black hole framework.

\begin{figure}
    \centering
     \includegraphics[scale=0.61]{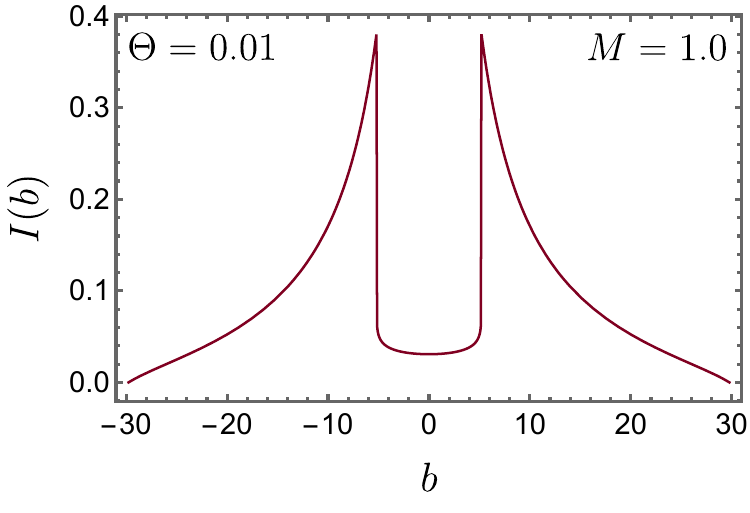}
    \includegraphics[scale=0.75]{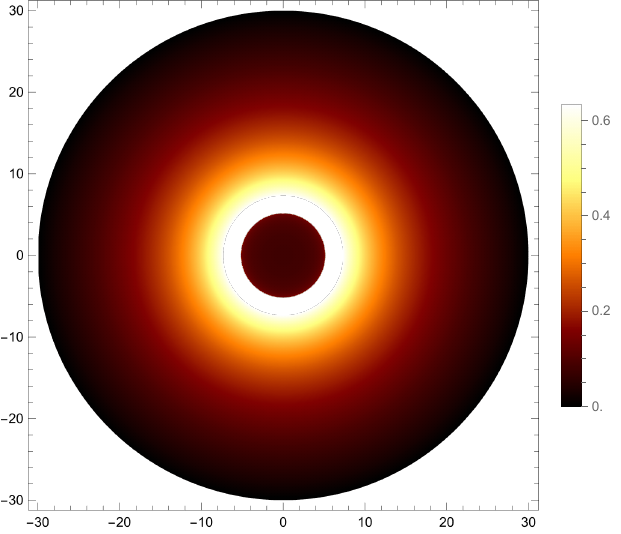}
     \includegraphics[scale=0.61]{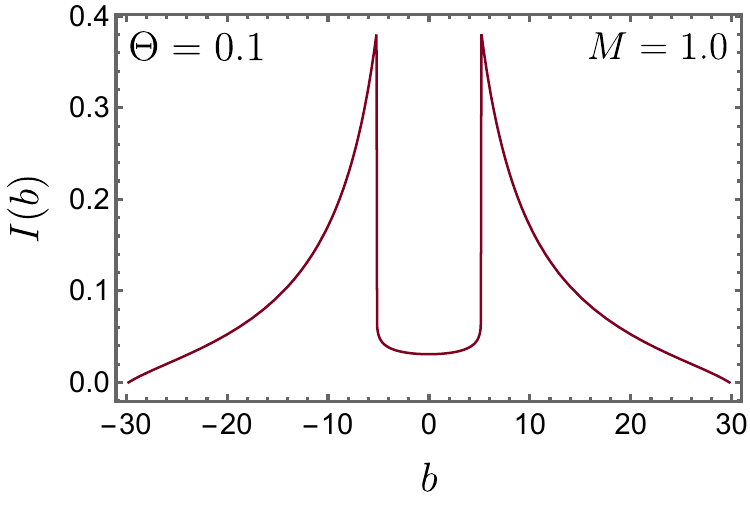}
    \includegraphics[scale=0.75]{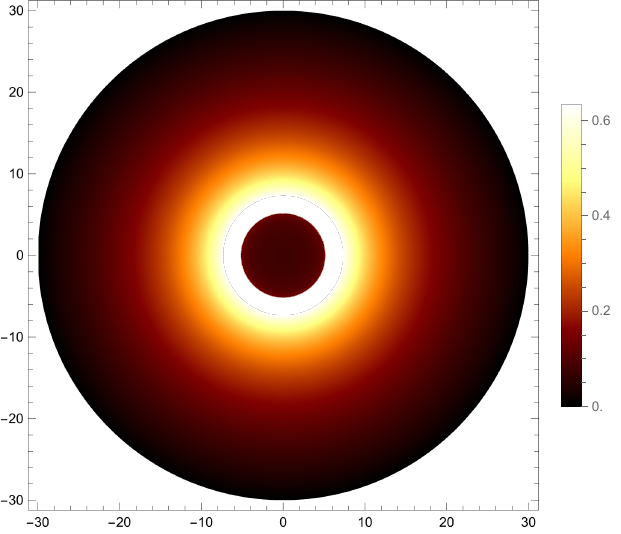}
      \includegraphics[scale=0.61]{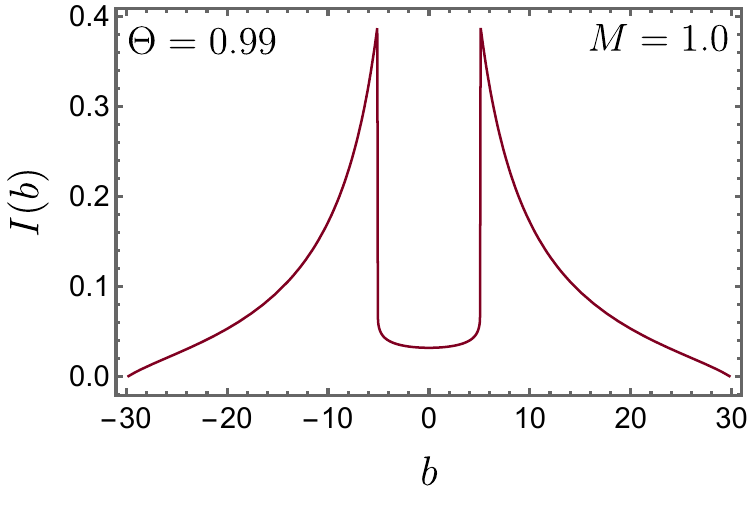}
    \includegraphics[scale=0.75]{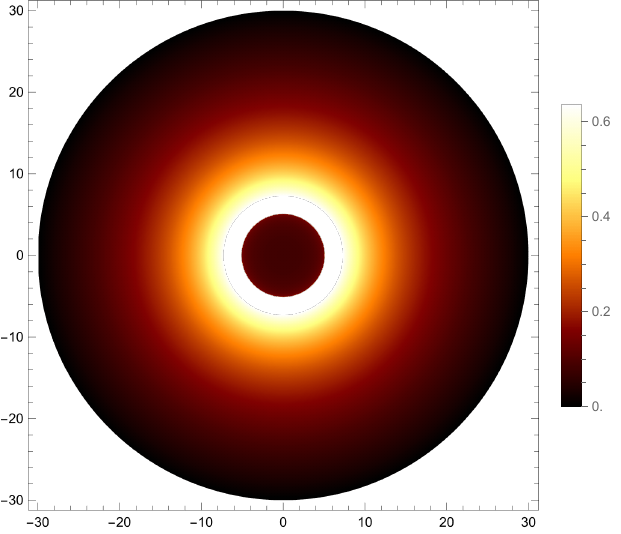}
    \caption{The shadows are shown for different values of the non--commutative parameter $\Theta$.}
    \label{accretionshadows}
\end{figure}

\begin{figure}
    \centering
     \includegraphics[height=6cm]{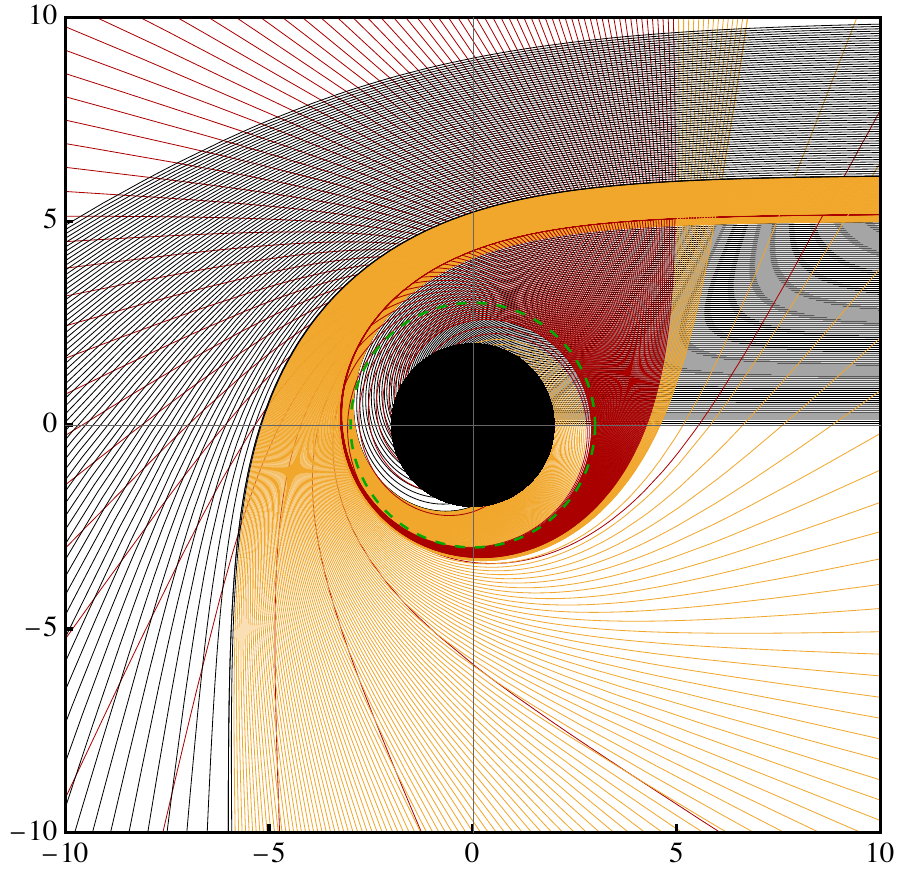}
    \includegraphics[height=6cm]{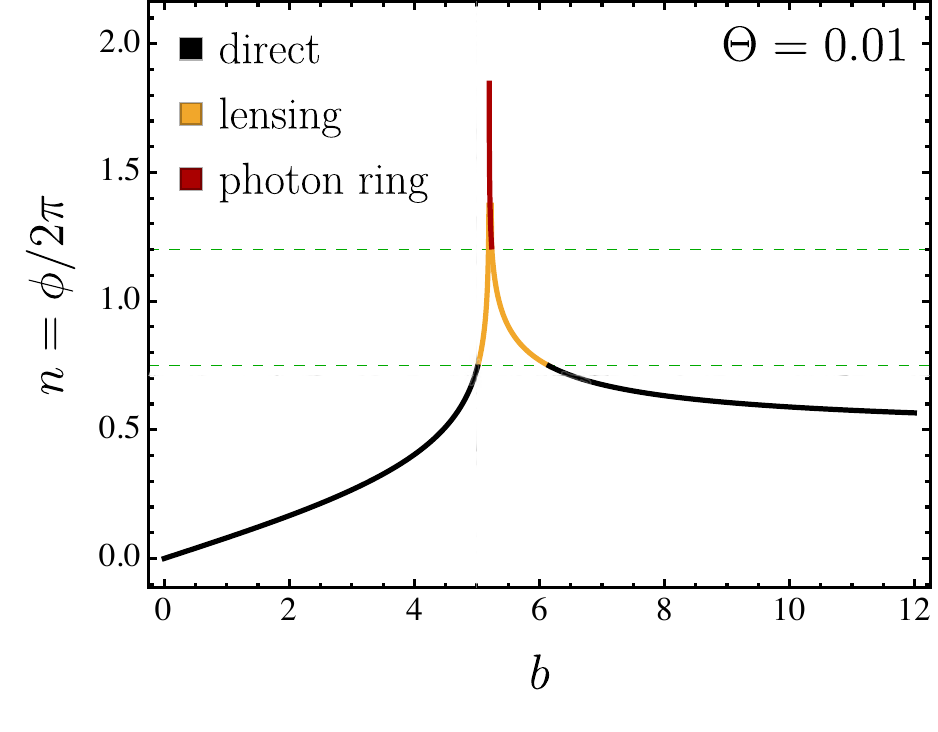}
   \includegraphics[height=6cm]{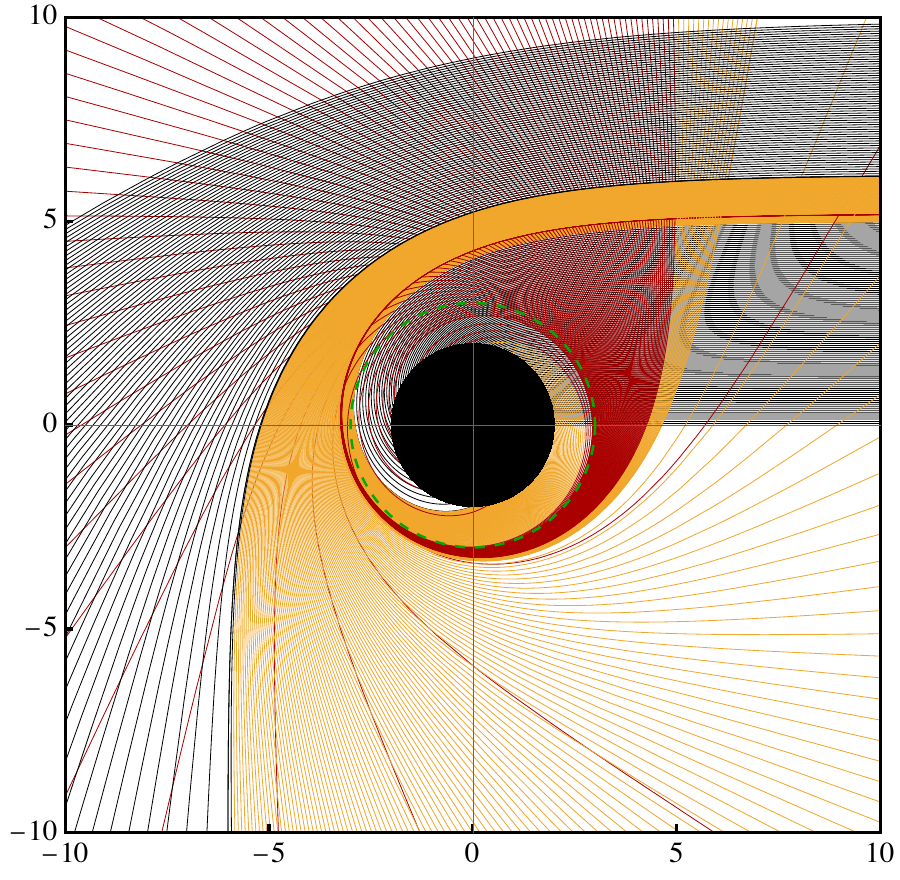}
    \includegraphics[height=6cm]{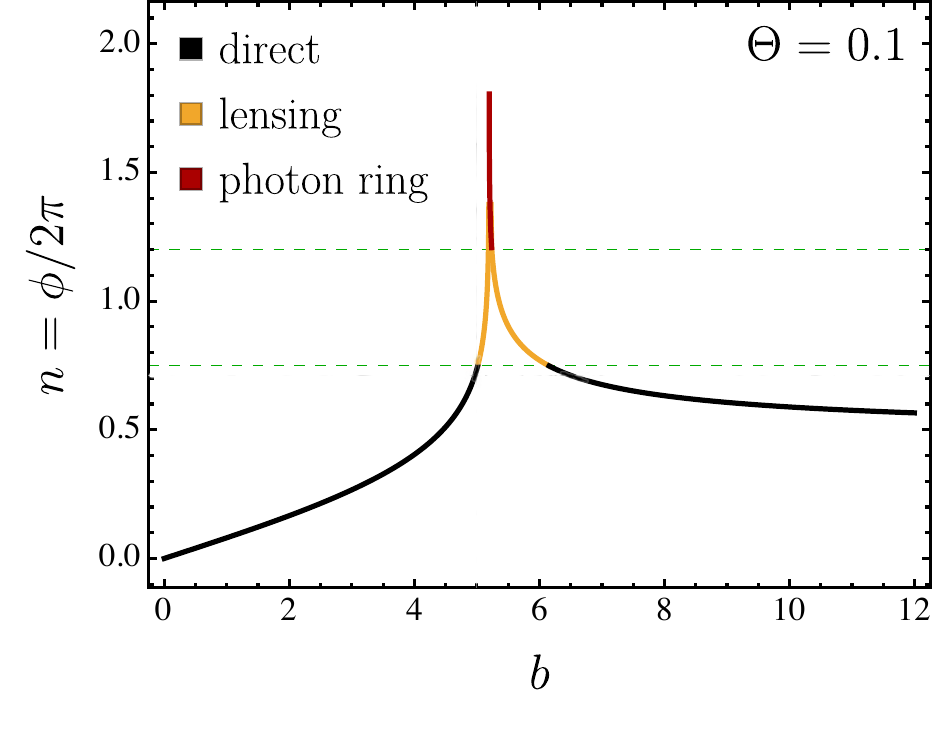}
    \includegraphics[height=6cm]{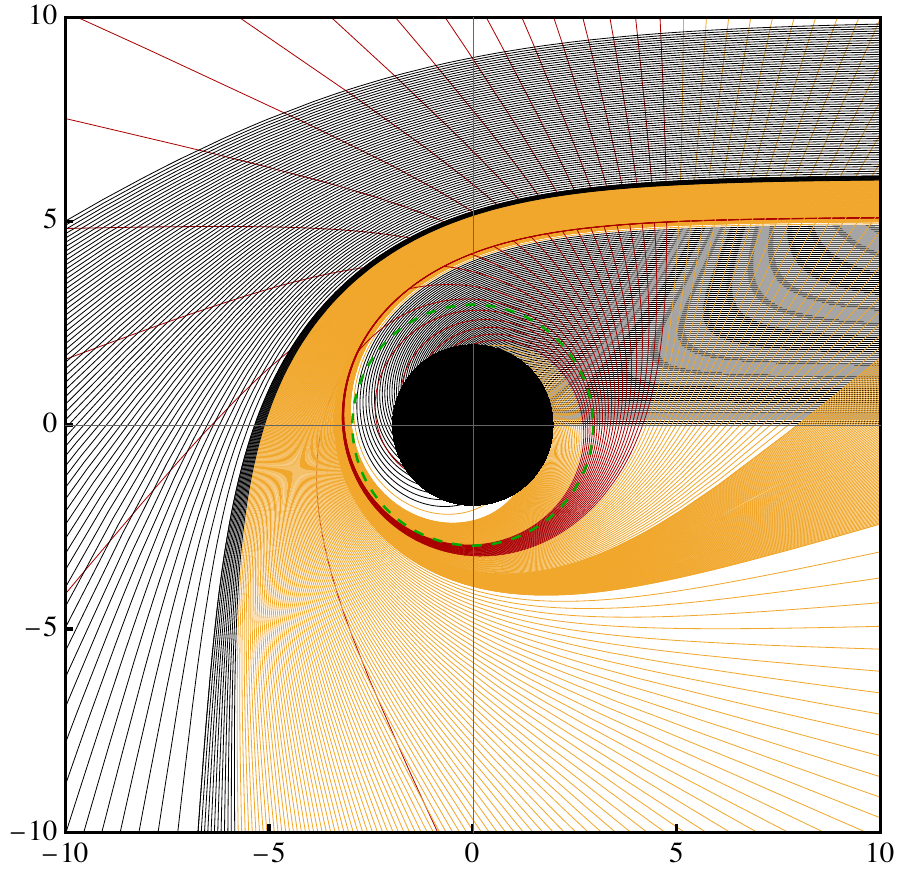}
    \includegraphics[height=6cm]{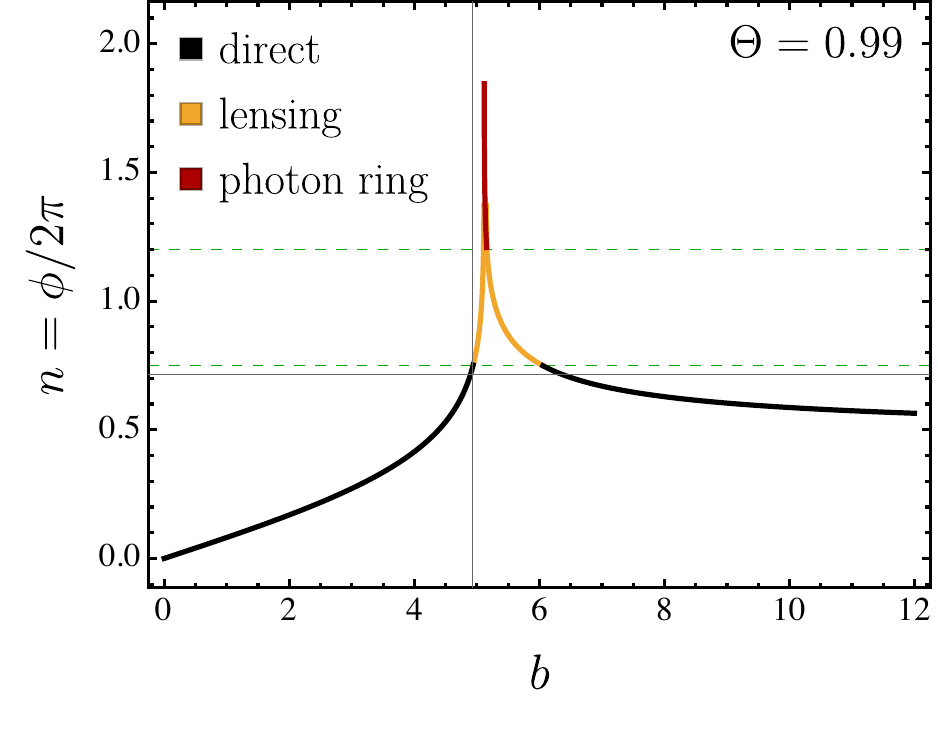}
    \caption{The regions linked to the black hole's direct emission, lensing rings, and photon rings are shown for various values of the non--commutative parameter \(\Theta\).}\label{accretion1}
\end{figure}

In Fig \ref{accretionshadows}, the total observed intensity $ I_{\text{obs}}$ of a black hole with non--commutativity, surrounded by an infalling accretion flow, is depicted as a function of the impact parameter $b$. The intensity displays a pronounced increase just before reaching its maximum value. The panels also display the two--dimensional shadows projected onto celestial coordinates for this configuration. Furthermore, as the black hole's parameter $\Theta$ increases, the observed intensity decreases, while the shadow size progressively expands. Furthermore, in Fig. \ref{accretion1}, the impact parameter regions associated with the black hole's direct emission, lensing rings, and photon rings are systematically studied for different values of the non--commutative parameter $\Theta$.

\begin{figure}[htp]
    \centering
\includegraphics[scale=0.51]{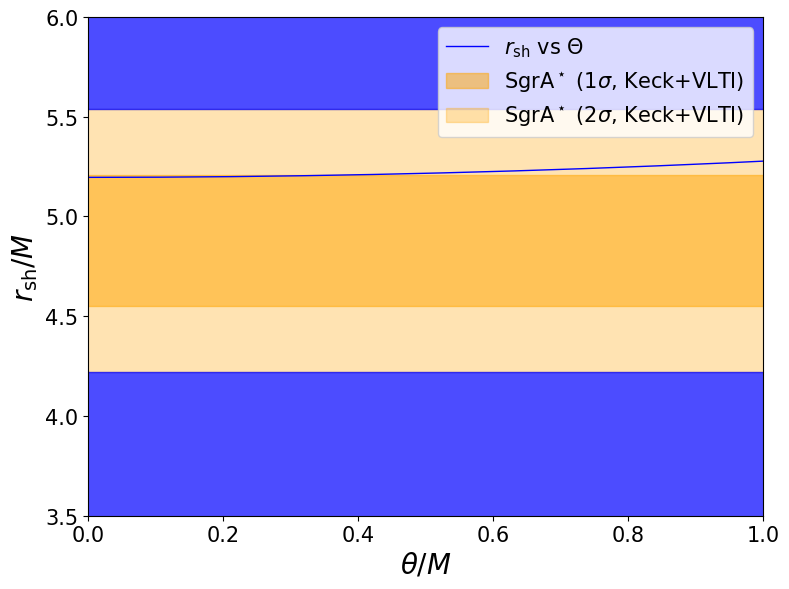}
    \caption{Analyze the variation of the normalized black hole shadow radius, \(r_{sh}/M\), as a function of \(\Theta/M\), while incorporating constraints on the parameter \(\beta\) based on Event Horizon Telescope (EHT) observations of Sgr A*.}
    \label{shadow}
\end{figure}

Figure \ref{shadow} illustrates the variation in the normalized black hole shadow radius, \(r_{sh}/M\), as a function of \(\Theta/M\). The results reveal that the shadow radius increases with higher values of \(\Theta/M\), consistent with the data presented in Table \ref{tab:photon_sphere_radii}. Furthermore, Figure \ref{shadow} incorporates constraints on the parameter \(\beta\) derived from the Event Horizon Telescope (EHT) observations of Sgr A*. Notably, the analysis at the \(68\%\) confidence level (C.L.) \cite{vagnozzi2022horizon} establishes an upper limit of \(\Theta/M \leq 0.316632\).

\begin{table}[]
\caption{The regions of the impact parameter associated with the black hole's direct emission, lensing rings, and photon rings are analyzed for varying values of the $\Theta$ parameter}
\label{tab:rings}
\begin{tabular}{|c|c|c|c|}
\hline
$\Theta$ & Photon ring ($n$\textgreater{}0.75)        & Lensed ring (0.75\textless{}$n$\textless{}1.25 )  & Direct emission ($n$\textless{}0.75)                                                     \\ \hline
0.01                   & 5.19615\textless $b$ \textless{}5.23615      & \begin{tabular}[c]{@{}c@{}}5.01615\textless{}$b$\textless{}5.19615;\\ 5.23615\textless{}$b$\textless{}6.15615\end{tabular}         & \begin{tabular}[c]{@{}c@{}}$b$\textless{}5.016615;\\ $b$\textgreater{}6.15615\end{tabular}     \\ \hline
0.1                   & 5.19534 \textless $b$ \textless{}5.23534 & \begin{tabular}[c]{@{}c@{}}5.015534\textless{}$b$\textless{}5.19534;\\ 5.23534\textless{}$b$\textless{}6.15534\end{tabular} & \begin{tabular}[c]{@{}c@{}}$b$\textless{}5.01534;\\ $b$\textgreater{}6.15534\end{tabular} \\ \hline
0.99                   & 5.11658 \textless $b$ \textless{}5.15658 & \begin{tabular}[c]{@{}c@{}}4.94658\textless{}$b$\textless{}5.11658;\\ 5.15658\textless{}$b$\textless{}6.05658\end{tabular} & \begin{tabular}[c]{@{}c@{}}$b$\textless{}4.94658;\\ $b$\textgreater{}6.05658\end{tabular} \\ \hline
\end{tabular}
\end{table}

\section{Lensing in the weak field regime}

This section addresses the application of the \textit{Gauss–Bonnet} theorem to calculate the weak deflection angle of a black hole. The analysis begins with the derivation of null geodesics by imposing the condition \(\mathrm{d}s^2 = 0\). Rearranging this condition leads to the following expression:
\begin{eqnarray}
\mathrm{d}t^2=\gamma_{ij}\mathrm{d}x^i \mathrm{d}x^j=\frac{1}{f_{\Theta}(r)^2}\mathrm{d}r^2+\frac{r^2}{f_{\Theta}(r)}\mathrm{d}\Omega^2.
\label{opmetric}
\end{eqnarray}
Here,  $f_{\Theta} = 1 - 2M_{\Theta}/r$ defines the metric function, while the indices $i$ and $j$ run from $1$ to $3$, and $\gamma_{ij}$ represents the components of the optical metric. To apply the \textit{Gauss--Bonnet} theorem, the Gaussian curvature must be determined, which is computed below
\ie
\begin{split}
\mathcal{K} & = \frac{R}{2}= \frac{f_{\Theta}(r)}{2} \frac{\mathrm{d}^{2}}{\mathrm{d} r^{2}}f_{\Theta}(r) -\frac{\left(\frac{\mathrm{d}}{\mathrm{d} r}f_{\Theta}(r)\right)^{2}}{4} \\
& =  \frac{3 M^2}{r^4} -\frac{2 M}{r^3} -\frac{3 \Theta ^2}{32 r^4} +\frac{\Theta ^2}{32 M r^3} +\frac{3 \Theta ^4}{4096 M^2 r^4} .
\end{split}
\fe
In this context, $\gamma$ represents the determinant of the optical metric $\gamma_{ij}$, while $R$ corresponds to the Ricci scalar. The surface area confined to the equatorial plane is expressed as \cite{Gibbons:2008rj}:
\begin{equation}
\mathrm{d}S=\sqrt\gamma \mathrm{d}r \mathrm{d} \phi= \frac{r}{f_{\Theta}(r)^{3/2}} \mathrm{d}r \mathrm{d}\phi =     \frac{r \, \mathrm{d}r \mathrm{d}\phi}{\left(1-\frac{2 \left( M - \frac{\Theta ^2}{64 M} \right)}{r}\right)^{3/2}}  .
\end{equation}

After establishing the necessary preliminaries, the deflection angle in the weak deflection limit is
\ie
\begin{split}
\label{lensing1}
& \alpha (b,\Theta)  = -\int\int\mathcal{K}\mathrm{d}S=-\int^{\pi}_0\int^{\infty}_{\Tilde{r}}\mathcal{K}\mathrm{d}S,
\end{split}
\fe
where $\Tilde{r}$, incorporating higher--order terms \cite{Jha:2024ltc}, is given by
\ie
u = \frac{1}{\Tilde{r}} = \frac{\sin\phi}{b} + \frac{M(1-\cos\phi)^2}{b^2}-\frac{M^2(60\phi\,
\cos\phi+3\sin3\phi-5\sin\phi)}{16b^3}.
\fe
Accordingly, after performing the integration, Eq. (\ref{lensing1}) becomes
\ie
\begin{split}
\alpha (b,\Theta) \simeq \, \frac{4 M}{b} -\frac{\Theta ^2}{4 b^3} + \frac{8 M^2}{b^3} + \frac{32 M^3}{3 b^3} -\frac{\Theta ^2}{16 b M} + \frac{136 M^4}{5 b^5} -\frac{\Theta ^2 M}{2 b^3} -\frac{17 \Theta ^2 M^2}{10 b^5}.
\end{split}
\fe

\begin{figure}
    \centering
     \includegraphics[scale=0.5]{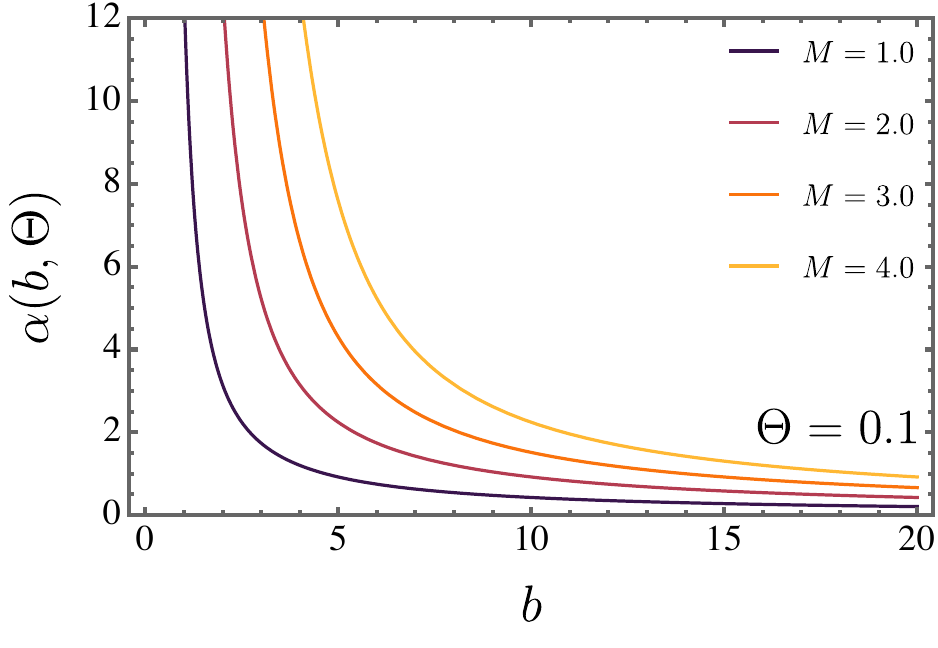}
    \includegraphics[scale=0.52]{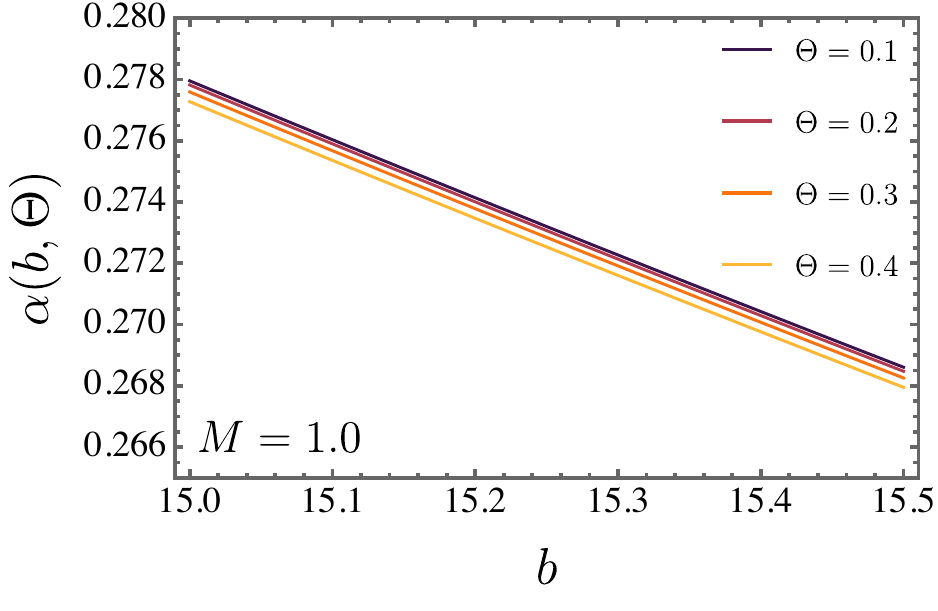}
     \includegraphics[scale=0.5]{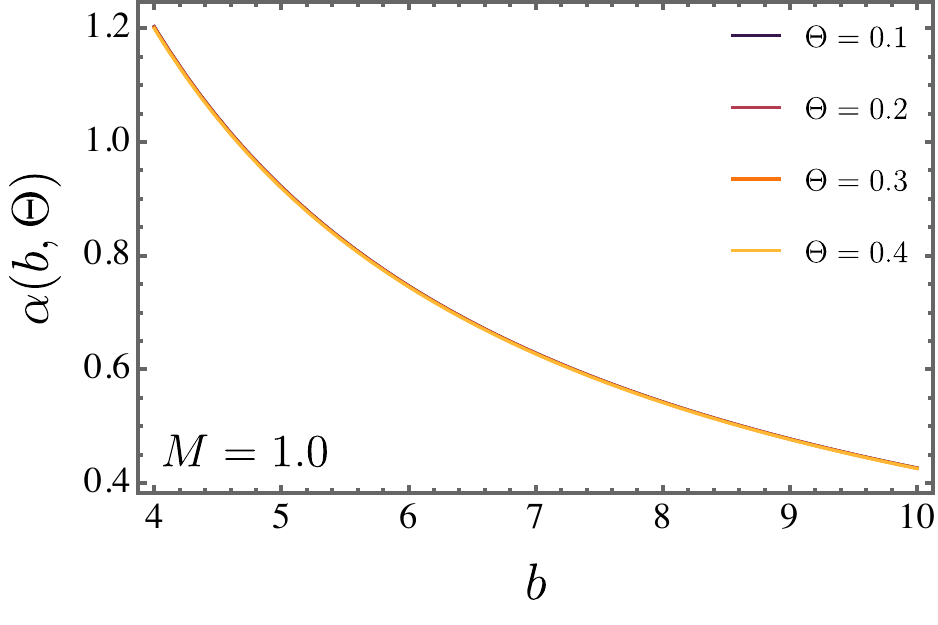}
    \caption{The deflection angle as a function of $b$ for different values of $M$ and $\Theta$.}
    \label{asdasd}
\end{figure}

The derivation of the above expression follows the same approximations as those used in Ref. \cite{Gibbons:2008rj}, specifically assuming  $b \gg 2M$ . Additionally, the non--commutative parameter $\Theta$ is included up to the second order. It is worth noting that the first four terms correspond to the Schwarzschild solution, expanded up to fourth order in $M$. To clarify Eq. (\ref{lensing1}), Fig. \ref{asdasd} is presented. The top--left panel illustrates that an increase in mass enhances the magnitude of the weak deflection angle. Similarly, the top--right and bottom panels indicate that higher charge values also result in an increase in $\alpha(b, \Theta)$.

\section{Lensing in the strong field regime}

This section focuses on deriving a mathematical expression for the deflection angle experienced by a photon traveling from infinity \((r \to \infty)\) as it passes near a gravitational source. This phenomenon, referred to as gravitational lensing, is analyzed in the strong field regime, where the photon’s trajectory comes very close to the massive object. Unlike the weak field regime, this scenario requires a more robust mathematical approach, which is outlined as follows.

To investigate the deflection angle in the strong field regime, we apply the framework introduced in \cite{tsukamoto2017deflection}. This method is specifically designed for spacetimes that are static, spherically symmetric, and asymptotically flat. The corresponding geometry is expressed through the following line element
\ie
\label{ssst}
\mathrm{d}s^{2} = - A(r) \mathrm{d}t^{2} + B(r) \mathrm{d}r^{2} + C(r)(\mathrm{d}\theta^2 + \sin^{2}\theta\mathrm{d}\phi^2),
\fe
In this setup, the metric functions satisfy the conditions \(\lim_{r \to \infty}A(r) = 1\), \(\lim_{r \to \infty}B(r) = 1\), and \(\lim_{r \to \infty}C(r) = r^2\). The symmetries of the spacetime give rise to two Killing vectors, \(\partial_t\) and \(\partial_\phi\), which correspond to the conservation of two quantities: the energy \(E\) and the angular momentum \(L\) \cite{Carroll:2004st}.

Using this approach, certain key quantities must be defined. The closest distance at which the photon approaches the gravitational object is represented by \( r_0 \), while the radius of the photon sphere, corresponding to the stable circular orbit of a photon, is denoted as \( r_m \). The strong field regime is characterized by the limit \( r_0 \to r_m \). Additionally, it is essential to define the impact parameter:
\ie
b \equiv \frac{L}{E} = \frac{C(r)\Dot{\phi}}{A(r)\Dot{t}}.
\fe

Owing to the symmetry of the specified metric, the geodesic equations governing photon trajectories simplify to the following form (refer to \cite{tsukamoto2017deflection} for a comprehensive explanation):
\ie
\left(  \frac{\mathrm{d}r}{\mathrm{d}\phi}     \right)^{2} = \frac{R(r)C(r)}{B(r)}.
\fe
Here, \( R(r) \equiv \frac{C(r)}{A(r)b^2} - 1 \), and the deflection angle of light, \(\alpha(r_0)\), is expressed as:
\ie
\alpha(r_{0}) = I(r_{0}) - \pi.
\fe
In this manner, $I(r_{0})$ is defined as shown below 
\ie
I(r_{0}) \equiv 2 \int^{\infty}_{r_{0}} \frac{\mathrm{d}r}{\sqrt{\frac{R(r)C(r)}{B(r)}}}.
\fe
Following the methodology proposed by Tsukamoto \cite{tsukamoto2017deflection}, we introduce a new variable defined as:
\ie
z \equiv 1 - \frac{r_{0}}{r},
\fe
so that the integral can be rewritten as
\ie
\label{int}I(r_{0}) = \int^{1}_{0} f(z,r_{0}) \mathrm{d}z,
\fe
where 
\ie
f(z,r_{0}) \equiv \frac{2r_{0}}{\sqrt{G(z,r_{0})}}, \,\,\,\,\,\,\,\, \text{and} \,\,\,\,\,\,\,\,  G(z,r_{0}) \equiv R \frac{C}{B}(1-z)^{4}.
\fe

The process continues by expanding \( G(z, r_0) \) as a power series in \( z \) and subsequently taking the limit \( r_0 \to r_m \), which corresponds to the strong field regime. Through further mathematical manipulation (as detailed in \cite{tsukamoto2017deflection}), it is evident that the integral \( I(r_0) \) exhibits a logarithmic divergence, necessitating a regularization procedure. To address this divergence, the integral \( I(r_0) \) is separated into two parts: a divergent component \( I_D(r_0) \) and a finite, regular component \( I_R(r_0) \). The specific steps for computing \( I_D(r_0) \) are not elaborated here, as they are thoroughly discussed in \cite{tsukamoto2017deflection}. The regularized portion of the integral \eqref{int}, expressed in terms of the impact parameter, is given by:
\ie
I_{R}(b) = \int^{1}_{0} f_{R}(z,b_{c})\mathrm{d}z + \mathcal{O}[(b-b_{c})\ln(b-b_{c})],
\fe
with 
\ie
b_{c}(r_{m}) \equiv \lim_{r_{0} \to r_{m}} \sqrt{\frac{C_{0}}{A_{0}}}.
\fe
The critical impact parameter is defined, with the subscript “\( m \)” indicating quantities evaluated at \( r = r_0 \). Additionally, \( f_R \equiv f(z, r_0) - f_D(z, r_0) \) represents the regularized function. In the strong field limit, the corresponding deflection angle reads
\ie
a(b) = - \Tilde{a} \ln \left[ \frac{b}{b_{c}}-1    \right] + \Tilde{b} + \mathcal{O}[(b-b_{c})\ln(b-b_{c})],
\label{deflections}
\fe
in which
\ie
\Tilde{a} = \sqrt{\frac{2 B_{m}A_{m}}{C^{\prime\prime}_{m}A_{m} - C_{m}A^{\prime\prime}_{m}}}, \,\,\,\,\,\,\,\, \text{and} \,\,\,\,\,\,\,\, \Tilde{b} = \Tilde{a} \ln\left[ r^{2}_{m}\left( \frac{C^{\prime\prime}_{m}}{C_{m}}  -  \frac{A^{\prime\prime}_{m}}{C_{m}} \right)   \right] + I_{R}(r_{m}) - \pi.
\fe
In the subsequent sections, this approach is applied to non--commutative black hole studied here to analyze the influence of $\Theta$ effects on the deflection angle.


\subsection{A non--commutative black hole via mass deformation}

This section focuses on analyzing the case where the cosmological constant is excluded, as the Tsukamoto method \cite{tsukamoto2017deflection} is specifically designed for spacetimes that are asymptotically flat. With the methodology established earlier, we now implement it for the metric under consideration, leading to the following results:
\ie
b_{c} = 3 \sqrt{3}  \left( M - \frac{\Theta ^2}{64 M} \right).
\fe
Moreover, the coefficients $\Tilde{a}$ and $\Tilde{b}$ are explicitly written as
\ie
\Tilde{a} = 1,
\fe
so that leading to
\ie
\begin{split}
& \Tilde{b} =  \ln 6
+ I_{R}(r_{m}) - \pi.
\end{split}
\fe
Also, $I_{R}(r_{m})$ is 
\ie
\begin{split}
 & I_{R}(r_{m}) =   \int_{0}^{1} \mathrm{d}z \left\{\frac{2 \sqrt{(3-2 z) z^2}-2 \sqrt{3} \sqrt{z^2}}{\sqrt{z^2} \sqrt{(3-2 z) z^2} \, \text{sgn}\left(\Theta ^2-64 M^2\right)} \right\} \\
 & = \ln \left[\frac{1}{36} \left(4 \sqrt{3}+7\right)\right] \text{sgn}\left(\Theta ^2-64 M^2\right).
\end{split}
\fe
Remarkably, this calculation is performed analytically. After that, the deflection angle can properly be addressed
\ie
\begin{split}
a(b) = &  -\ln  \left[  \frac{b}{3 \sqrt{3}  \left( M - \frac{\Theta ^2}{64 M} \right)}   - 1 \right] + \ln \left[\frac{1}{36} \left(4 \sqrt{3}+7\right)\right] \text{sgn}\left(\Theta ^2-64 M^2\right) - \pi \\
& + \mathcal{O}\left\{ \left[ b - 3 \sqrt{3}  \left( M - \frac{\Theta ^2}{64 M} \right) \right] \ln \left[ b- 3 \sqrt{3}  \left( M - \frac{\Theta ^2}{64 M} \right)\right] \right\}.
\end{split}
\fe

\begin{figure}
    \centering
     \includegraphics[scale=0.5]{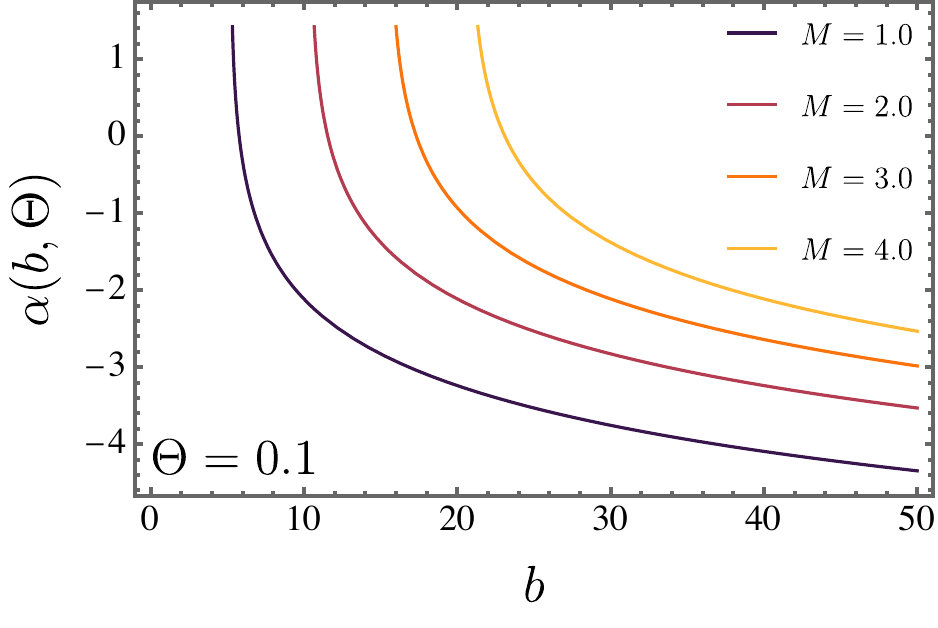}
    \includegraphics[scale=0.525]{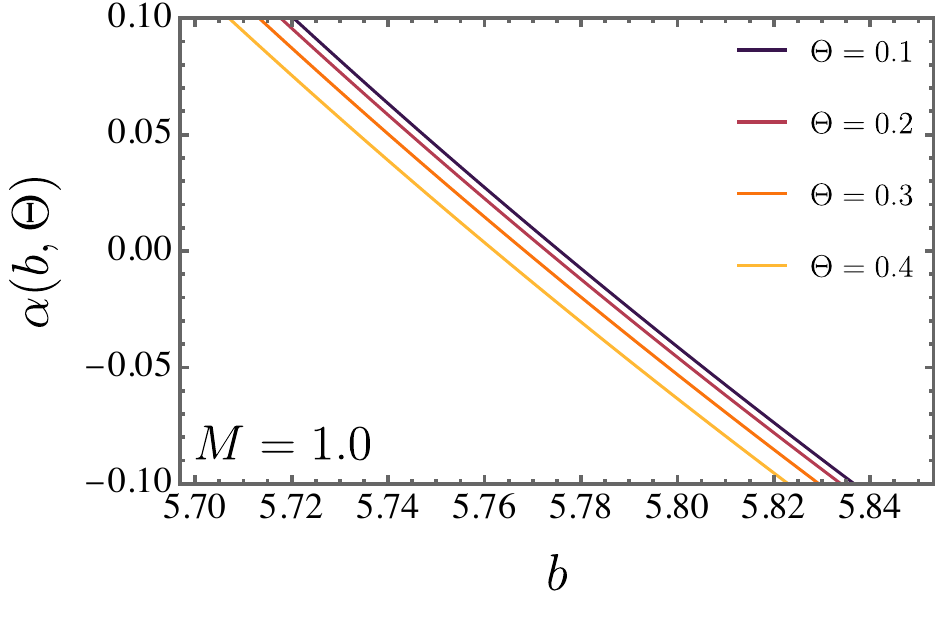}
     \includegraphics[scale=0.5]{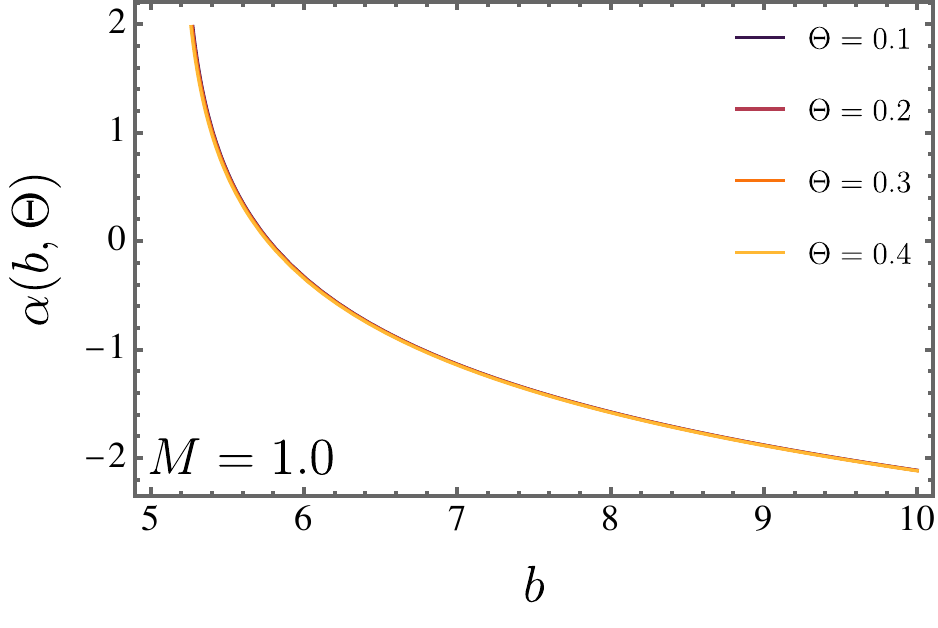}
    \caption{The deflection angle as a function of $b$ for different values of $M$ and $\Theta$.}
    \label{sdjidjsi}
\end{figure}

To aid comprehension, Fig. \ref{sdjidjsi} illustrates how the deflection angle varies with $b$ under different system configurations. In general lines, the findings from the strong deflection limit are consistent with those from the weak field approximation addressed earlier. Specifically, the top--left panel shows that a higher mass increases the deflection angle, while the top--right and bottom panels indicate that greater charge values also result in a larger $a(b, \Theta)$.

\section{Lensing equations and observables}

This section focuses on analyzing the parameters that govern the bending of light in the strong gravitational field surrounding the black hole. Light emitted from the source \( S \) (indicated by the red point) is deflected due to the gravitational field of the non--commutative black hole at \( L \) (orange point), eventually reaching the observer \( O \) (purple point). The observer perceives the resulting image at \( I \) (blue point) and the big black disk represents the black hole itself. The black hole being studied is represented by a large black dot. The angular position of the source is identified as \(\beta\), while \(\theta\) denotes the angular position of the observed image. The path deviation of the light, represented by \( a \), quantifies the alteration in its trajectory caused by the gravitational influence.

This study utilizes the framework proposed in \cite{030, bozza2001strong}, which considers a scenario where the source (\( S \)) is nearly aligned with the lens (\( L \)). Such an arrangement is crucial for the generation of relativistic images. In this context, the lens equation, establishing the connection between the angular positions \(\theta\) and \(\beta\), is formulated as $
\beta = \theta - \frac{D_{LS}}{D_{OS}} \Delta a_{n}$.  
The term \(\Delta a_{n}\) denotes the deflection angle, accounting for all photon loops before reaching the observer, and is defined as \(\Delta a_{n} = a - 2n\pi\). Using this approach, the impact parameter is approximated as \(\Tilde{b} \approx \theta D_{OL}\), allowing the angular deviation to be expressed in the following form: $
\label{angle}
a(\theta)=- \Tilde{a}\ln\left(\frac{\theta D_{OL}}{b_c}-1\right)+\Tilde{b}$.

To calculate \(\Delta a_n\), the deflection angle \(a(\theta)\) is expanded around \(\theta = \theta^0_n\), where \(\theta^0_n\) satisfies the condition \(\alpha(\theta^0_n) = 2n\pi\). This expansion results in:  
\ie
\Delta a_n = \frac{\partial a}{\partial \theta} \Bigg|_{\theta=\theta^0_n} (\theta - \theta^0_n).
\fe  
The approximate angular position \(\theta^0_n\) is expressed as:  
\ie
\theta^0_n = \frac{b_c}{D_{OL}}(1 + e_n), \quad \text{with} \quad e_n = e^{\Tilde{b} - 2n\pi}.
\fe  
Substituting this into the deflection angle leads to the following expression:  
\ie
\Delta a_n = -\frac{\Tilde{a} D_{OL}}{b_c e_n} (\theta - \theta^0_n).
\fe 
This relationship is then incorporated into the lens equation to determine the angular position of the \(n\)-th relativistic image:  
\ie
\theta_n \simeq \theta^0_n + \frac{b_c e_n}{\Tilde{a}} \frac{D_{OS}}{D_{OL}D_{LS}} (\beta - \theta^0_n).
\fe  
While the deflection of light preserves the surface brightness of the source, the lensing effect modifies the apparent solid angle, affecting the perceived brightness. The magnification of the image, \(\mu_n\), is defined as:  
\ie
\mu_n = \left| \frac{\beta}{\theta} \frac{\partial \beta}{\partial \theta} \bigg|_{\theta^0_n} \right|^{-1}.
\fe  
By using the relation \(\Delta a_n = -\frac{\Tilde{a} D_{OL}}{b_c e_n} (\theta - \theta^0_n)\), the magnification factor can be written as:  
\ie
\mu_n = \frac{e_n(1 + e_n)}{\Tilde{a} \beta} \frac{D_{OS}}{D_{LS}} \left(\frac{b_c}{D_{OL}}\right)^2.
\fe

As \( n \) increases, the magnification factor \( \mu_n \) also grows, indicating that the brightness of the primary relativistic image, \(\theta_1\), is substantially greater than that of the subsequent images. Despite this, the overall luminosity remains limited, primarily due to the dependence on the term \( \left(\frac{b_c}{D_{OL}}\right)^2 \). Importantly, the magnification diverges as \( \beta \to 0 \), underscoring that near-perfect alignment between the source and the lens significantly enhances the visibility of relativistic images, which aligns with expectations.

Additionally, the impact parameter can be directly associated with the angular position \(\theta_\infty\), representing the asymptotic limit of the remaining relativistic images. This relationship is expressed as \cite{bozza2001strong}:  
\ie
b_c = D_{OL} \theta_\infty.
\fe  

In line with Bozza's approach presented in \cite{bozza2001strong}, the outermost image \(\theta_1\) is treated as a distinct feature, while the remaining images are grouped under \(\theta_\infty\). To explore the properties of these images, Bozza introduced the following observables:  
\ie
s=\theta_{1}-\theta_{\infty}= \theta_{\infty} e^{\frac{\Tilde{b}-2\pi}{\Tilde{a}}} \ , \quad
\tilde{r} =  \frac{\mu_{1}}{\sum\limits_{n=2}^{\infty} \mu_{n} }= e^{\frac{2\pi}{\Tilde{a}}} .
\fe.

The angular separation, denoted by $s$, and the flux ratio, represented by $\tilde{r}$, describe key observational quantities. The flux ratio quantifies the relative contribution of the brightest image compared to the combined flux from all other images. By rearranging these relations, it becomes possible to derive the coefficients of the expansion. To illustrate these concepts, the next subsection will focus on a specific astrophysical scenario, providing calculations for these observables and examining the impact of non--commutativity on the derived parameters.

\begin{figure}
    \centering
     \includegraphics[scale=0.5]{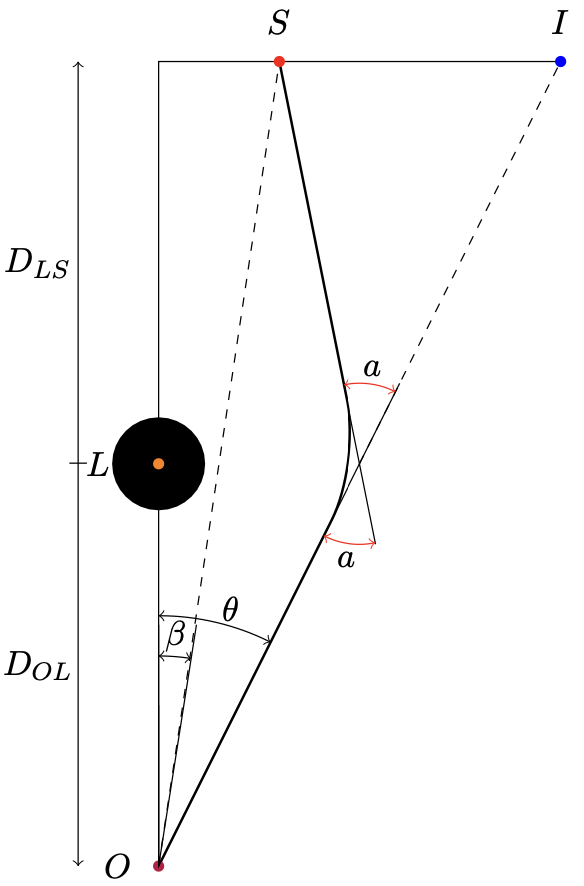}
    \caption{Representation of the gravitational lensing. The light emitted from the source \( S \) (red point) is bent as it travels toward the observer \( O \) (purple point), influenced by the presence of a compact object positioned at \( L \) (orange point). The observer \( O \) perceives an image \( I \) (blue point). \( D_{OL} \) represents the distance between the lens \( L \) and the observer \( O \), while \( D_{LS} \) denotes the distance from the source's projection to the lens along the optical axis. The big black dot represents the black hole under consideration }
    \label{obs1}
\end{figure}


\subsection{Gravitational lensing analysis utilizing Sagittarius \( A^{*} \) data}

Stellar dynamics observations strongly point to the presence of a compact and mysterious object at the center of our galaxy. This object, identified as the supermassive black hole Sagittarius (Sgr) \( A^{*} \), is estimated to have a mass of \( 4.4 \times 10^6 M_{\odot} \) \cite{genzel2010galactic}. To investigate its properties further, the study utilizes parameter $\Theta$, which plays a important role in analyzing and characterizing the behavior of relevant observables.

To examine the observables, a distance of \( D_{OL} = 8.5 \, \text{Kpc} \) is considered \cite{genzel2010galactic}, along with a non--commutative parameter value of \( \Theta \sim 1.235 \times 10^{-35} \, \text{m} \) as suggested in the literature \cite{kobakhidze2016constraining,touati2024quantum}. Using \( b_{c} = 3 \sqrt{3}  \left( M - \frac{\Theta ^2}{64 M} \right) \), the angular size is calculated as \( \theta_{\infty} \approx 25.24 \, \mu\text{arcsecs} + \mathcal{O}(\Theta^{2}) \), where \( \mathcal{O}(\Theta^{2}) \) accounts for second--order contributions of the non--commutative parameter.

Gravitational lensing phenomena have been explored within the framework of the charged Simpson--Visser solution \cite{zhang2022gravitational}. The angular size \( \theta_{\infty} \) for Sagittarius \( A^{*} \) is reported to vary between \( 20.7 \, \mu\text{as} \) and \( 26.6 \, \mu\text{as} \), while the deviation \( \delta \theta_{\infty} \) ranges from \(-6.0 \, \mu\text{as}\) to \( 0 \, \mu\text{as}\). Both parameters, \( \theta_{\infty} \) and \( \delta \theta_{\infty} \), exhibit a decrease as the charge \( q \) increases \cite{zhang2022gravitational}. Although \( \theta_{\infty} \) falls within the detection capabilities of the Event Horizon Telescope (EHT), the small deviation \( \delta \theta_{\infty} \), which can reach up to \( 6 \, \mu\text{as} \), exceeds the EHT's current observational precision. Consequently, differentiating between the black--bounce--Reissner--Nordström spacetime and the Schwarzschild black hole based on \( \delta \theta_{\infty} \) remains unattainable at this time.


\section{Time delay}

The theoretical framework underpinning our investigation is presented in this section. To compute the time delay experienced by light in a gravitational field, we analyze the null geodesics within a spherically symmetric spacetime. The metric governing this spacetime is expressed as
\ie
\mathrm{d}\tau^{2} = f_{\Theta}(r)\mathrm{d}t^{2} -\frac{1}{f_{\Theta}(r)}\mathrm{d}r^{2} 
	-r^{2}(\mathrm{d}\theta^{2}+\sin^{2}\theta \mathrm{d}\phi^{2}),
\fe
where the geodesic equations yield conserved quantities associated with the motion of particles. These quantities include the angular momentum \(L = r^{2}\sin^{2}\theta \frac{\mathrm{d}\phi}{\mathrm{d}\lambda}\), the energy \(E = f_{\Theta}(r)\frac{\mathrm{d}t}{\mathrm{d}\lambda}\), and the norm of the four-velocity \(\mathcal{L}\), which can be expressed as
\ie
\mathcal{L} = g_{\mu\nu}\mathrm{d}x^{\mu}\mathrm{d}x^{\nu} = f_{\Theta}(r) \bigg( \frac{\mathrm{d}t}{\mathrm{d}\lambda} \bigg)^2
		- \frac{1}{f_{\Theta}(r)} \bigg( \frac{\mathrm{d}r}{\mathrm{d}\lambda} \bigg)^{2}
		- r^{2} \bigg( \frac{\mathrm{d}\theta}{\mathrm{d}\lambda} \bigg)^{2} 
		- r^{2}\sin^{2}\theta \bigg( \frac{\mathrm{d}\phi}{\mathrm{d}\lambda} \bigg)^{2}.
\fe
Here, \(\lambda\) serves as the affine parameter, \(L\) represents the conserved angular momentum, and \(E^2/2\) characterizes the conserved energy of the particle along its trajectory. By restricting the motion to the equatorial plane (\(\theta = \pi/2\)), the differential equations governing the geodesics simplify to
\ie
		\frac{1}{2} \bigg( \frac{\mathrm{d}r}{\mathrm{d}\lambda} \bigg)^{2} + \frac{1}{2} f_{\Theta}(r) \bigg[ \frac{L^{2}}{r^{2}} + \mathcal{L} \bigg]
		= \frac{1}{2} \bigg( \frac{\mathrm{d}r}{\mathrm{d}\lambda} \bigg)^{2} + V(r)
		= \frac{1}{2}E^{2}.
\fe
Within this framework, the effective potential governing particle dynamics in a spherically symmetric gravitational field is described by  
\ie
V(r) = \frac{f_{\Theta}(r)}{2} \left[ \frac{L^2}{r^2} + \mathcal{L} \right],
\fe 
where $b = |L/E|$ denotes the impact parameter. For massless particles traveling along null geodesics, the term \( \mathcal{L} \) vanishes. By concentrating on the motion of photons, the expression simplifies, leading to:  
\ie
\frac{\mathrm{d}r}{\mathrm{d}t} = \frac{\mathrm{d}r}{\mathrm{d}\lambda}  \frac{\mathrm{d}\lambda}{\mathrm{d}t}
= \pm f_{\Theta}(r) \sqrt{1 - b^{2}\frac{f_{\Theta}(r)}{r^{2}}}.
\fe
For a massless photon, where \( \mathcal{L} = 0 \), and using the expression \( E = f_{\Theta}(r) \frac{\mathrm{d}t}{\mathrm{d}\lambda} \), the \( \pm \) signs represent distinct phases of the motion. Initially, as the photon moves from its source at \( r_{\text{S}} \), the radial distance \( r \) decreases steadily, reaching a minimum value at \( r = r_0 \), which marks the closest approach to the black hole. Once this turning point is crossed, the radial distance begins to increase as the photon continues its path outward. Based on this interpretation, the following relationships are derived:
\ie
\frac{\mathrm{d}r}{\mathrm{d}t} = - f_{\Theta}(r) \sqrt{1 - b^{2} \frac{f_{\Theta}(r)}{r^{2}}} < 0.
\fe

As a photon travels from its initial location at \( r = r_{\text{S}} \) toward the turning point at \( r = r_{0} \), the radial coordinate \( r \) decreases steadily, indicating its approach to the black hole. Upon reaching \( r = r_{0} \), the point of closest approach, the direction of motion changes, and \( r \) begins to increase as the photon moves away from the black hole. This progression characterizes the photon’s trajectory, governed by the equations describing its geodesic motion:
\ie
\frac{\mathrm{d}r}{\mathrm{d}t} = f_{\Theta}(r) \sqrt{1 - b^{2} \frac{f_{\Theta}(r)}{r^{2}}} > 0,  
\fe
Considering the segment of the photon's path from the turning point at \( r = r_{0} \) to the observer's location at \( r = r_{\text{O}} \), the radial coordinate \( r \) increases as the photon moves outward. In the context of gravitational lensing, where the light source is positioned at \( r = r_{\text{S}} \) and the observer at \( r = r_{\text{O}} \), the time delay experienced by the light as it traverses the gravitational field can be expressed as follows, based on the formulation provided in \cite{qiao2024time}:
\ie
\begin{split}
\label{asdasdddd}
 \Delta T & = T - T_{0} \\
 & = -\int_{r_{\text{S}}}^{r_{0}} \frac{\mathrm{d}r}{f_{\Theta}(r)\sqrt{1-\frac{b^{2} f_{\Theta}(r)}{r^{2}}}}
	      + \int_{r_{0}}^{r_{\text{O}}} \frac{\mathrm{d}r}{f_{\Theta}(r)\sqrt{1-\frac{b^{2} f_{\Theta}(r)}{r^{2}}}}
	      - T_{0}
	      \\
	& = \int_{r_{0}}^{r_{\text{S}}} \frac{\mathrm{d}r}{f_{\Theta}(r)\sqrt{1-\frac{b^{2} f_{\Theta}(r)}{r^{2}}}}
          + \int_{r_{0}}^{r_{\text{O}}} \frac{\mathrm{d}r}{f_{\Theta}(r)\sqrt{1-\frac{b^{2} f_{\Theta}(r)}{r^{2}}}}
          - \sqrt{r_{\text{S}}^{2}-r_{0}^{2}} - \sqrt{r_{\text{O}}^{2}-r_{0}^{2}}.
\end{split}
\fe
In the absence of a gravitational field, the propagation time for light is given by \( T_{0} = \sqrt{r_{\text{S}}^{2} - r_{0}^{2}} + \sqrt{r_{\text{O}}^{2} - r_{0}^{2}} \). When the gravitational effects are included, the time delay, \( \Delta T \), grows continuously with increasing distances \( r_{\text{S}} \) (light source) and \( r_{\text{O}} \) (observer). By considering the $\Theta$ and $b$ small, Eq. (\ref{asdasdddd}) reads
\ie
\begin{split}
\Delta T = &\, \frac{1}{2} b^2 \left(\frac{1}{r_{0}-r_{\text{O}}}+\frac{1}{r_{0}- r_{\text{S}}}\right)+\frac{1}{16} \Theta ^2 \left(\frac{1}{-2 M - r_{0} + r_{\text{O}}}+\frac{1}{-2 M-r_{0} + r_{\text{S}}}\right)-2 r_{0} + r_{O} + r_{\text{S}} \\ 
& +\frac{b^4 \left(128 \left(\frac{1}{r_{0}^3-r_{\text{O}}^3}+\frac{1}{r_{0}^3 - r_{\text{S}}^3}\right)-\frac{3 \left(64 M^2-\Theta ^2\right) \left(2 r_{0}^4- r_{\text{O}}^4-r_{\text{S}}^4\right)}{M \left(r_{0}^4- r_{\text{O}}^4\right) \left(r_{0}^4 - r_{\text{S}}^4\right)}\right)}{1024} \\ 
& +\frac{(8 M-\Theta ) (\Theta +8 M) (\ln (-2 M - r_{0}+r_{\text{O}}) + \ln (-2 M - r_{0} + r_{\text{S}}))}{32 M} \\ & - \sqrt{r_{\text{S}}^{2}-r_{0}^{2}} 
 - \sqrt{r_{\text{O}}^{2}-r_{0}^{2}}.
\end{split}
\fe

\begin{figure}
    \centering
     \includegraphics[scale=0.615]{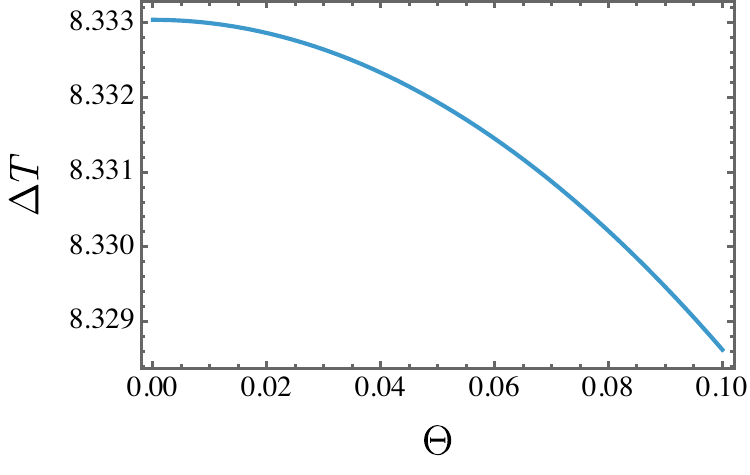}
     \includegraphics[scale=0.6]{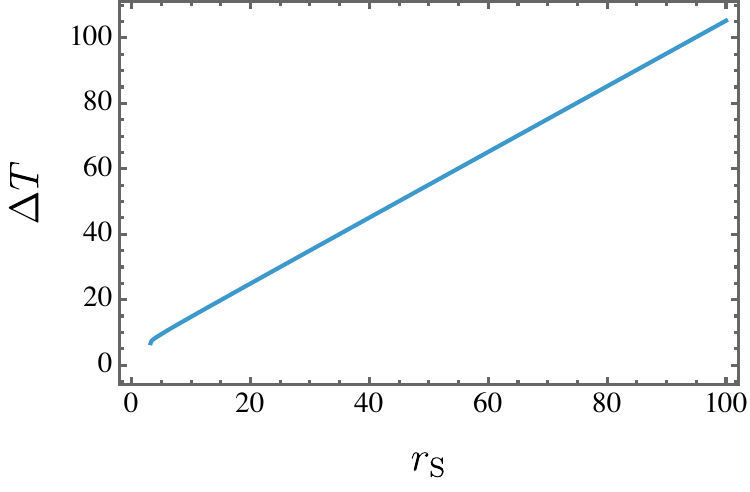}
    \caption{The time delay \(\Delta T\) is evaluated as a function of $\Theta$ (left panel) and \(r_{\text{S}}\) (right panel) for the parameters \( \Theta = 0.1 \) (only for the right panel), $M= 0.1$, $r_{0} = 3$, $r_{\text{O}} = 10$, $b=0.1$ and $r_{\text{S}} = 4$ (only for the left panel).}
    \label{timedelay}
\end{figure}

To illustrate this behavior, Fig. \ref{timedelay} depicts the time delay \(\Delta T\) as a function of \( r_{\text{S}} \), calculated for the parameter values \( \Theta = 0.1 \) (only for the right panel), $M= 0.1$, $r_{0} = 3$, $r_{\text{O}} = 10$, $b=0.1$ and $r_{\text{S}} = 4$ (only for the left panel).

\section{The energy deposition rate by the neutrino annihilation process}

We begin by analyzing the energy deposition within the spacetime influenced by a non--commutative black hole with mass deformation. The energy deposition rate per unit time and volume, resulting from the neutrino pair annihilation process, is expressed as \cite{Salmonson:1999es}:
\begin{align}
\dfrac{\mathrm{d}E(\mathbf{r})}{\mathrm{d}t\mathrm{d}V}=2KG_{F}^{2}F(r)\iint
n(\varepsilon_{\nu})n(\varepsilon_{\overline{\nu}})
(\varepsilon_{\nu}+\varepsilon_{\overline{\nu}})
\varepsilon_{\nu}^{3}\varepsilon_{\overline{\nu}}^{3}
\mathrm{d}\varepsilon_{\nu}\mathrm{d}\varepsilon_{\overline{\nu}}
\end{align}
where
\begin{align}
K=\dfrac{1}{6\pi}(1\pm4\sin^{2}\theta_{W}+8\sin^{4}\theta_{W}).
\end{align}
Using the Weinberg angle \(\sin^{2}\theta_{W} = 0.23\), the expressions for various neutrino pairs are given as \cite{Salmonson:1999es}:
\begin{align}
K(\nu_{\mu},\overline{\nu}_{\mu})=K(\nu_{\tau},\overline{\nu}_{\tau})
=\dfrac{1}{6\pi}\left(1-4\sin^{2}\theta_{W}+8\sin^{4}\theta_{W}\right)
\end{align}
and
\begin{align}
K(\nu_{e},\overline{\nu}_{e})
=\dfrac{1}{6\pi}\left(1+4\sin^{2}\theta_{W}+8\sin^{4}\theta_{W}\right)
\end{align}
respectively \cite{Salmonson:1999es}. It is important to note that the Fermi constant is \( G_{F} = 5.29 \times 10^{-44} \, \text{cm}^{2} \, \text{MeV}^{-2} \). Accordingly, the angular integration factor is expressed as follows \cite{Salmonson:1999es}:
\begin{align}
F(r)&=\iint\left(1-\bm{\Omega_{\nu}}\cdot\bm{\Omega_{\overline{\nu}}}\right)^{2}
\mathrm{d}\Omega_{\nu}\mathrm{d}\Omega_{\overline{\nu}}\notag\\
&=\dfrac{2\pi^{2}}{3}(1-x)^{4}\left(x^{2}+4x+5\right)\hspace{1cm}
\end{align}
with
\begin{align}
x=\sin\theta_{r}.
\end{align}

The angle \(\theta_{r}\) represents the inclination between the particle's trajectory and the tangent vector to a circular orbit at a radius \(r\). For a given type of neutrino or antineutrino, \(\Omega_{\nu} (\Omega_{\overline{\nu}})\) denotes the unit direction vector, while \(\mathrm{d}\Omega_{\nu} (\mathrm{d}\Omega_{\overline{\nu}})\) corresponds to the differential solid angle. At temperature \(T\), the number densities of neutrinos and antineutrinos in phase space, \(n(\varepsilon_{\nu})\) and \(n(\varepsilon_{\overline{\nu}})\), respectively, follow the Fermi--Dirac distribution \cite{Salmonson:1999es} 
\begin{align}
n(\varepsilon_{\nu})=\dfrac{2}{h^{3}}\dfrac{1}{\exp\left({\dfrac{\varepsilon_{\nu}}{kT}}\right)+1}.
\end{align}
Here, \( h \) denotes Planck's constant and \( k \) represents Boltzmann's constant. By integrating Eq. (5), the energy deposition per unit time and unit volume is determined as \cite{Salmonson:1999es}
\begin{align}
\dfrac{\mathrm{d}E}{\mathrm{d}t\mathrm{d}V}=\dfrac{21\zeta(5)\pi^{4}}{h^{6}}KG_{F}^{2}F(r)(kT)^{9}.
\end{align}
Deriving the expression for \(\frac{\mathrm{d}E}{\mathrm{d}t\mathrm{d}V}\) is crucial for advancing studies on energy conversion rates in various compact objects \cite{Salmonson:1999es}. This expression depends on spatial position and incorporates the temperature \(T = T(r)\), referred to as the local temperature \cite{Salmonson:1999es}.

The local temperature \( T(\mathbf{r}) \), as measured by a local observer, is defined by the relation \( T(\mathbf{r})\sqrt{g^{\Theta}_{tt}(\mathbf{r})} = \text{constant} \), where \( g^{\Theta}_{tt} \) is a metric component of the spacetime \cite{Salmonson:1999es}. The temperature of neutrinos at the neutrinosphere is described as \cite{Salmonson:1999es}
\begin{align}
T(r)\sqrt{g^{\Theta}_{tt}(r)}=T(R)\sqrt{g^{\Theta}_{tt}(R)}
\end{align}
where \( R \) represents the radius of the gravitational source. For future calculations, it is convenient to substitute the local temperature \( T(r) \) using the relation in identity (13). The luminosity, incorporating the effects of redshift, which reads \cite{Salmonson:1999es}
\begin{align}
L_{\infty} = g^{\Theta}_{tt}(R_{0})L(R_{0})
\end{align}
in which the luminosity for a single neutrino species at the neutrinosphere is \cite{Salmonson:1999es}
\begin{align}
L(R)=4\pi R_{0}^{2}\dfrac{7}{4}\dfrac{ac}{4}T^{4}(R).
\end{align}

Here, \( a \) represents the radiation constant, and \( c \) is the speed of light in a vacuum. To express the temperature in terms of the observer's position, we have \cite{Salmonson:1999es}:
\begin{align}
\dfrac{\mathrm{d}E(\mathbf{r})}{\mathrm{d}t\mathrm{d}V}=\dfrac{21\zeta(5)\pi^{4}}{h^{6}}
KG_{F}^{2}k^{9}\left(\dfrac{7}{4}\pi ac\right)^{-\frac{9}{4}}L_{\infty}^{\frac{9}{4}}F(r)
\dfrac{\left[g^{\Theta}_{tt}(R)\right]^{\frac{9}{4}}}{\left[g^{\Theta}_{tt}(r)\right]^{\frac{9}{4}}}
R_{0}^{-\frac{9}{2}},
\end{align}
where, \(\zeta(s)\) denotes the Riemann zeta function, defined for \(s > 1\) as the infinite series:
\ie
\zeta(s) = \sum_{n=1}^{\infty} \frac{1}{n^s}.
\fe
In addition to the radial coordinate, the metric components at the massive source's surface contribute to the expression for the energy deposition rate per unit time and unit volume. The radiation energy power in the gravitational field can be determined by integrating the deposition energy density over time. To compute the angular integration \( F(r) \), it is necessary to further analyze the variable \( x \) introduced previously. In this manner, the null geodesic equations in the spacetime of a spherically symmetric gravitational source are solved \cite{Salmonson:1999es}, as demonstrated in \cite{Lambiase:2020iul,Shi:2023kid}:
\begin{align}
x^{2}&=\sin^{2}\theta_{r}|_{\theta_{R}=0}\notag\\
&=1-\dfrac{R^{2}}{r^{2}}\dfrac{f_{\Theta}(r)}{f_{\Theta}(R)}.\hspace{0.5cm}
\end{align}
Here, \( f_{\Theta}(r) = g^{\Theta}_{tt}(r) \). Relating the variable \( x = \sin\theta|_{\theta_{R}} = 0 \) to the surrounding structure of the gravitational source is insightful for understanding the environment \cite{Salmonson:1999es}. The angular integration factor becomes dependent on the metric. By applying this relationship, we can integrate the rate of energy deposition per unit time and unit volume over the spherically symmetric volume encompassing the gravitational source \cite{Lambiase:2020iul,Shi:2023kid}
\begin{align}
\dot{Q}&=\dfrac{\mathrm{d}E}{\sqrt{g^{\Theta}_{tt}}\mathrm{d}t}\notag\hspace{10cm}\\
&=\dfrac{84\zeta(5)\pi^{5}}{h^{6}}KG_{F}^{2}k^{9}
\left(\dfrac{7}{4}\pi ac\right)^{-\frac{9}{4}}
L_{\infty}^{\frac{9}{4}}\left[g^{\Theta}_{tt}(R)\right]^{\frac{9}{4}}
R^{-\frac{9}{2}}\int_{R_{0}}^{\infty}\dfrac{r^{2}\sqrt{- g^{\Theta}_{rr}(r)}F(r)}
{g^{\Theta}_{tt}(r)}\mathrm{d}r.
\end{align}

In this context, \( g^{\Theta}_{rr}(r) = -\frac{1}{f_{\Theta}(r)} \). The quantity \(\dot{Q}\) represents the total energy converted from neutrinos into electron-positron pairs per unit time at a given radius \cite{Salmonson:1999es}. When \(\dot{Q}\) reaches extremely high values, this conversion process can lead to explosive outcomes. Comparing the energy deposition rate, with Newtonian quantities is essential for further analysis \cite{Salmonson:1999es,Lambiase:2020iul,Shi:2023kid}
\begin{align}
\dfrac{\dot{Q}}{\dot{Q}_{Newt}}=3\left[g^{\Theta}_{tt}(R)\right]^{\frac{9}{4}}
\int_{1}^{\infty}(x-1)^{4}\left(x^{2}+4x+5\right)\dfrac{y^{2}\sqrt{-g^{\Theta}_{rr}(Ry)}}
{g^{\Theta}_{tt}(Ry)^{\frac{9}{2}}}\mathrm{d}y.
\end{align}
Using the dimensionless variable \( y = \frac{r}{R} \) and the metric components \( g^{\Theta}_{tt}(r) \) and \( g^{\Theta}_{rr}(r) \) from Eq. (1), we can express \(\frac{\mathrm{d}\dot{Q}}{\mathrm{d}r}\) as a function of the radial coordinate \( r \). This formulation highlights the variations and enhancements in the energy deposition rate with respect to \( r \)
\begin{align}
\dfrac{\mathrm{d}\dot{Q}}{\mathrm{d}r}&=4\pi\left(\dfrac{\mathrm{d}E}{\mathrm{d}t\mathrm{d}V}\right)\sqrt{-g^{\Theta}_{rr}(r)}r^{2}\hspace{9.5cm}\notag\\
&=\dfrac{168\zeta(5)\pi^{7}}{3h^{6}}KG_{F}^{2}k^{9}
\left(\dfrac{7}{4}\pi ac\right)^{-\frac{9}{4}}
L_{\infty}^{\frac{9}{4}}\notag\\
&\quad\times(x-1)^{4}\left(x^{2}+4x+5\right)
\left[\dfrac{g^{\Theta}_{tt}(R)}{g^{\Theta}_{tt}(r)}\right]^{\frac{9}{4}}R^{-\frac{5}{2}}
\sqrt{-g^{\Theta}_{rr}(r)}\left(\dfrac{r}{R}\right)^{2}.
\end{align}

The derivative \(\frac{\mathrm{d}\dot{Q}}{\mathrm{d}r}\) depends on the radial coordinate, originating from the center of the gravitational source, and incorporates the metric functions. Understanding how the structural properties of compact objects in the framework of asymptotic safety influence neutrino annihilation is essential, particularly in identifying conditions under which such annihilation could lead to gamma--ray bursts. After some algebraic manipulations, we have
\cite{Salmonson:1999es,Lambiase:2020iul,Shi:2023kid}:
\begin{align}
\dfrac{\dot{Q}}{\dot{Q}_{Newt}}=3\left[f_{\Theta}(R)\right]^{\frac{9}{4}}\int_{1}^{\infty}
(x-1)^{4}\left(x^{2}+4x+5\right)\dfrac{y^{2}}{\left[f(Ry)\right]^{5}}\mathrm{d}y,
\end{align}
where
\begin{align}
f_{\Theta}(R)&= 1 - 2M_{\Theta}/R,\\
f_{\Theta}(Ry)&=1-\dfrac{2M_{\Theta}}{R}\dfrac{1}{y}.
\end{align}
And, therefore, we write
\begin{align}
x^{2}=1-\dfrac{1}{y^{2}}\dfrac{1-\dfrac{2M_{\Theta}}{R}\dfrac{1}{y}
}{1-\dfrac{2M_{\Theta}}{R}}.
\end{align}

\begin{figure}[t]
\includegraphics[width=15cm]{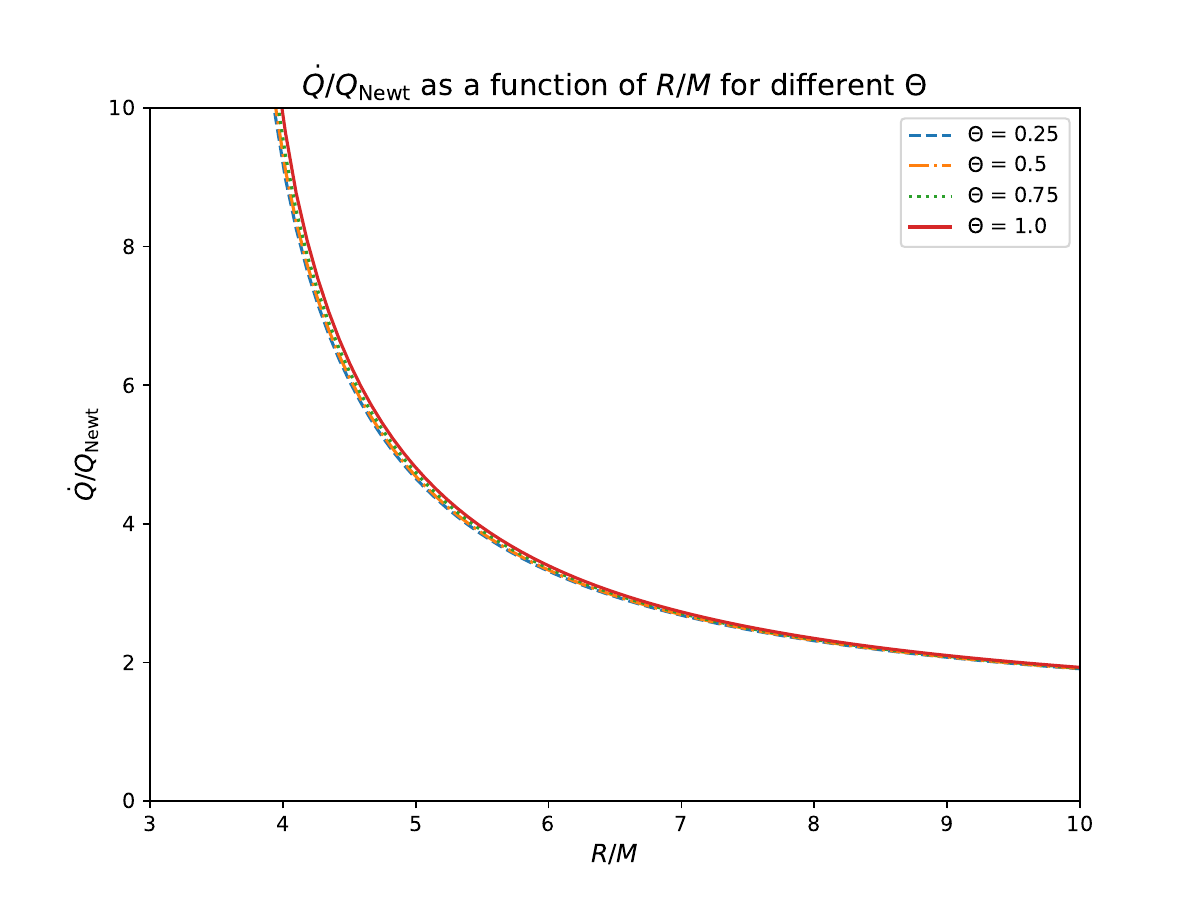}
\caption{The solid, dotted and dashed curves of the ratio $\dfrac{\dot{Q}}{\dot{Q_{Newt}}}$ as functions of the ratio $\dfrac{R}{M}$ for the parameter $\Theta=0.25, 0.5, 0.75, 1$ respectively.}
\label{energydeposition}
\end{figure}

To enhance the interpretation of our findings, we present Fig. \ref{energydeposition}, which illustrates $\dot{Q}/\dot{Q}_{Newt}$ as a function of $R/M_{\Theta}$ for various values of the non--commutative parameter \(\Theta\). The plot reveals that \(\dot{Q}/\dot{Q}_{Newt}\) increases as \(\Theta\) becomes larger.


\section{Neutrino oscillation phase and probability}

The spacetime geometry for a static and spherically symmetric configuration is described by the metric:
\ie
\label{maininin}
\mathrm{d}s^{2} = f_{\Theta}(r)\mathrm{d}t^2-\dfrac{\mathrm{d}r^2}{f_{\Theta}(r)}-r^2\left(\mathrm{d}\theta^2+\sin^2\theta\mathrm{d}\varphi^2\right).
\fe
Within a spherically symmetric spacetime, as described by the metric (\ref{maininin}), the motion of neutrinos in the \(k\)--th eigenstate is governed by the following Lagrangian \cite{neu18}:
\ie
\mathcal{L}
= \dfrac{1}{2}m_{k} f_{\Theta}(r) \left(\dfrac{\mathrm{d}t}{\mathrm{d}\tau}\right)^2-\dfrac{1}{2}\dfrac{m_{k}}{f_{\Theta}(r)}\left(\dfrac{\mathrm{d}r}{\mathrm{d}\tau}\right)^2
-\dfrac{1}{2}m_{k}r^2\left(\dfrac{\mathrm{d}\theta}{\mathrm{d}\tau}\right)^2-\dfrac{1}{2}m_{k}r^2\sin^2\theta\left(\dfrac{\mathrm{d}\varphi}{\mathrm{d}\tau}\right)^2.
\fe

The proper time \(\tau\) and \(m_{k}\), representing the mass of the \(k\)--th eigenstate, define the canonical conjugate momentum for the coordinate \(x^{\mu}\) as \(p_{\mu} = \frac{\partial\mathcal{L}}{\partial\left(\frac{\mathrm{d}x}{\mathrm{d}\tau}\right)}\). When the particle's motion is confined to the equatorial plane (\(\theta = \frac{\pi}{2}\)), the nonzero components of the momentum are determined as follows \cite{neu60,Shi:2024flw}:
\ie
p^{(k)t} = m_{k}f_{\Theta}(r)\dfrac{\mathrm{d}t}{\mathrm{d}\tau} = E_{k}, \quad
p^{(k)r} = \dfrac{m_{k}}{f_{\Theta}(r)}\dfrac{\mathrm{d}r}{\mathrm{d}\tau}, \quad
p^{(k)\varphi} = m_{k}r^2\dfrac{\mathrm{d}\varphi}{\mathrm{d}\tau}=J_{k},
\fe
in which \(k\)--th eigenstate's mass adheres to the mass--shell condition \cite{neu54,neu55}:
\ie
m_{k}^2 = g_{\mu\nu}p^{(k)\mu}p^{(k)\nu}.
\fe

Neutrino flavor oscillations in curved spacetime have been explored using the plane wave approximation, especially in scenarios involving weak gravitational fields \cite{neu53,neu54}. In weak interactions, neutrinos are identified and detected based on their flavor eigenstates, as shown in \cite{neu61,neu62,Shi:2024flw,neu63}:
\ie
\ket{\nu_{\alpha}}=\sum_{i=1}^{3}U_{\alpha i}^{*}\ket{\nu_{i}}.
\fe

In this context, \(\alpha = e, \mu, \nu\) represents the three neutrino flavors, while the mass eigenstates are denoted by \(\ket{\nu_{i}}\). The leptonic mixing matrix \(U\), a \(3 \times 3\) unitary matrix, establishes the relationship between flavor eigenstates and mass eigenstates \cite{neu41}. Neutrino mass eigenstates and their propagation between two spacetime points can be described using wave functions. For convenience, the coordinates \(\left(t_{S},\bm{x}_{S}\right)\) and \(\left(t_{D},\bm{x}_{D}\right)\) are assigned to the source (\(S\)) and the detector (\(D\)), respectively. The wave function for such propagation is expressed as:
\ie
\ket{\nu_{i}\left(t_{D},\bm{x}_{D}\right)}=\exp\left(-\mathrm{i}\Phi_{i}\right)\ket{\nu_{i}\left(t_{S},\bm{x}_{S}\right)},
\fe
so that the phase reads 
\ie
\Phi_{i}=\int_{\left(t_{S},\bm{x}_{S}\right)}^{\left(t_{D},\bm{x}_{D}\right)}g_{\mu\nu}p^{(i)\mu}\mathrm{d}x^{\nu}.
\fe
We now revisit the phenomenon of flavor oscillation occurring during neutrino propagation from its generation point at the source to its detection at the detector. The probability of a neutrino flavor transition \(\nu_{\alpha} \to \nu_{\beta}\) at the detection location is determined as
\ie
P_{\alpha\beta}
= |\left\langle \nu_{\beta}|\nu_{\alpha}\left(t_D, \bm{x}_D\right)\right\rangle|^2 
=\sum_{i,j=1}^3U_{\beta i}U_{\beta j}^*U_{\alpha j}U_{\alpha i}^*\exp[-\mathrm{i}(\Phi_{i}-\Phi_{j})].
\fe

The motion of neutrinos confined to the equatorial plane (\(\theta = \frac{\pi}{2}\)) under the influence of a non--commutative black hole's gravitational field is analyzed. Therefore, the phase can be written below
\ie
\begin{split}
\label{Pgefhi}
\Phi_{k} &= \int_{\left(t_{S},\bm{x}_{S}\right)}^{\left(t_D, \bm{x}_D\right)}g_{\mu\nu}p^{(k)\mu}\mathrm{d}x^{\nu} = \int_{\left(t_{S},\bm{x}_{S}\right)}^{\left(t_D, \bm{x}_D\right)}\left[E_k\mathrm{d}t-p^{(k)r}\mathrm{d}r-J_k\mathrm{d}\varphi\right] 
= \pm\dfrac{m_k^2}{2E_0}\int_{r_S}^{r_D}\left\{1-\dfrac{b^2}{r^2}\left[f_{\Theta}(r)\right]\right\}^{-\frac{1}{2}}\mathrm{d}r \\
&\approx \pm\dfrac{m_k^2}{2E_0}\Biggl\{\left[\sqrt{r_D^2-b^2}-\sqrt{r_S^2-b^2}\right]
+  \left( M - \frac{\Theta ^2}{64 M} \right) \left[\dfrac{r_D}{\sqrt{r_D^2-b^2}}-\dfrac{r_S}{\sqrt{r_S^2-b^2}}\right] \Biggl\}.
\end{split}
\fe

Here, \( E_0 = \sqrt{E_k^2 - m_k^2} \) represents the average energy of relativistic neutrinos emitted from the source, and \( b \) denotes the impact parameter \cite{neu18}. As neutrinos propagate, their trajectory includes a closest approach point at \( r = r_0 \). Within the weak field approximation, the minimum distance \( r_0 \) is determined as a solution to the orbital equation governing neutrino motion
\begin{align}
\label{r0}
r_0 \simeq b -  \left( M - \frac{\Theta ^2}{64 M} \right).
\end{align}

The phase acquired by neutrinos during their propagation ---from the source, through the point of closest approach, to the detector --- is calculated using Eq. (\ref{Pgefhi}) along with Eq. (\ref{r0})
\ie
\label{pphhiii}
\begin{split}
&\Phi_k\left(r_S\to r_0\to r_D\right)\\
&\simeq \dfrac{m_k^2}{2E_0}
\biggl\{
\left[\sqrt{r_S^2-b^2}+\sqrt{r_D^2-b^2}\right] \\
& + \left( M - \frac{\Theta ^2}{64 M} \right)\biggl[\dfrac{b}{\sqrt{r_S^2-b^2}}
+\dfrac{b}{\sqrt{r_D^2-b^2}}
+\dfrac{\sqrt{r_S-b}}{\sqrt{r_S+b}}
+\dfrac{\sqrt{r_D-b}}{\sqrt{r_D+b}}\biggl]\biggr\}.
\end{split}
\fe

Expanding Eq. (\ref{pphhiii}) up to the order of \(\frac{b^2}{r_{S,D}^2}\) under the condition \(b \ll r_{S,D}\), we obtain
\ie
\begin{split}
\Phi_k
& \simeq \dfrac{m_k^2}{2E_0}
\left\{\left(r_S+r_D\right) \left[\left(1-\dfrac{b^2}{2r_Sr_D}\right)
+\dfrac{2  \left( M - \frac{\Theta ^2}{64 M} \right)}{r_S+r_D}\right] \right\}.
\end{split}
\fe

Gravitational lensing effects on neutrinos emerge during their propagation. To thoroughly analyze the neutrino flavor oscillation probability addressed previously in the vicinity of the black hole, it is essential to calculate the phase difference along the various possible paths \cite{Shi:2024flw}
\ie
\Delta\Phi_{ij}^{pq}
= \Phi_i^{p}-\Phi_j^{q}\notag\\
 = \left(\Delta m_{ij}^2 A_{pq}+\Delta b_{pq}^2 B_{ij}\right),
\fe
with
\ie
\begin{split}
\Delta m_{ij}^2 &= m_i^2-m_j^2,\quad
\Delta b_{pq}^2 = b_p^2-b_q^2,\\
A_{pq} &= \dfrac{r_S + r_D}{2E_0}\left\{1+\dfrac{2 \left( M - \frac{\Theta ^2}{64 M} \right)}{r_S+r_D}-\dfrac{\sum b_{pq}^2}{4r_S \, r_D}\right\},\\
B_{ij} &= -\dfrac{\sum m_{ij}^2}{8E_0}\left(\dfrac{1}{r_S}+\dfrac{1}{r_D}\right),\\
\sum b_{pq}^2 &= b_p^2+b_q^2,\quad
\sum m_{ij}^2 = m_i^2+m_j^2 .
\end{split}
\fe

To represent the phases corresponding to different neutrino paths, upper indices such as \(\Phi_{i}^{p}\) are introduced to specify the trajectory, with each path characterized by its impact parameter \(b_{p}\). The phase difference in the neutrino transition probability for various paths around a non--commutative black hole depends on the individual neutrino masses \(m_{i}\), the squared mass differences \(\Delta m_{ij}^{2}\), and the properties of the gravitational source. In addition, when the non--commutative parameter \(\Theta\) is set to zero, the phase difference reduces to the results reported in Ref. \cite{neu53}. The coefficient \(B_{ij}\) depends on the neutrino masses, while the modification to \(A_{pq}\) caused by non--commutativity influences both the phase and the oscillation amplitude. The coefficients \(A_{pq}\) and \(B_{ij}\) remain symmetric when their respective lower indices are swapped.


\section{Neutrino gravitational lensing}

Within the gravitational influence of a massive source, neutrinos can propagate along nonradial paths, and gravitational lensing may occur between the neutrino source and the detector \cite{neu54}. This phenomenon enables neutrinos taking different trajectories to reach the detector \(D\). As a result, it becomes necessary to reformulate the neutrino flavor eigenstate as \cite{Shi:2024flw,neu56,neu62,neu63,neu64,neu65}
\ie
|\nu_{\alpha}(t_{D},\textbf{x}_{D})\rangle=N\sum_{i}U_{\alpha i}^{\ast}
\sum_{p}\exp(-i\Phi_{i}^{p})|\nu_{i}(t_{S}, \textbf{x}_{S})\rangle,
\fe
with \( p \) denoting the path index. Considering that nearly all neutrinos converge at the detector, the probability of flavor transition \(\nu_{\alpha} \rightarrow \nu_{\beta}\) at the detection point is \cite{Shi:2024flw,neu56,neu62,neu63,neu64,neu65}
\ie
\label{nasndkas}
\mathcal{P}_{\alpha\beta}^{\mathrm{lens}}=|\langle\nu_{\beta}|\nu_{\alpha}(t_{D}, \textbf{x}_D)\rangle|^{2}\hspace{2cm}\notag\\
=|N|^{2}\sum_{i, j}U_{\beta i}U_{\beta j}^{\ast}U_{\alpha j}U_{\alpha j}^{\ast}\sum_{p, q}\exp(\Delta\Phi_{ij}^{pq}),
\fe
so that the normalization constant can be expressed as
\ie
|N|^{2} = \Biggl[\sum_{i}|U_{\alpha i}|^{2}\sum_{p,q}\exp(-i\Delta\Phi_{ij}^{pq})\Biggl]^{-1}.
\fe

Considering the phase difference \(\Delta\Phi_{ij}^{pq}\) discussed in the expressions, the neutrino oscillation likelihood in the context of neutrino lensing is influenced by several factors. These include the individual neutrino masses, the squared mass differences, and the structural characteristics of the black hole, as described in Eq. (\ref{nasndkas}). This behavior is analogous to that observed near spherically symmetric sources, such as the Schwarzschild black hole \cite{neu53}.

We focus on examining the probability of neutrino oscillations under the influence of gravitational lensing, emphasizing the role of the non--commutative parameter \(\Theta\). Within the framework where non--commutativity serves as a lens for two--flavor neutrinos, we investigate the transition probabilities for \(\nu_{\alpha} \to \nu_{\beta}\) at the detection point. The probability function is obtained in the weak--field approximation, considering the geometry defined by the source, lens, and detector \cite{neu53,neu54,neu55,Shi:2024flw,neu65}
\ie
\begin{split}
\label{asdPdadbd2}
\mathcal{P}_{\alpha\beta}^{\mathrm{lens}}
&= \left|N\right|^2\Biggl\{2\sum_i\left|U_{\beta i}\right|^2\left|U_{\alpha i}\right|^2\left[1+\cos (\Delta b_{12}^2B_{ii})\right]\\
&\quad + \sum_{i\neq j}U_{\beta i}U_{\beta j}^*U_{\alpha j}U_{\alpha i}^*\left[\exp\left(-\mathrm{i} \,\Delta m_{ij}^2 A_{11}'\right)+\exp\left(-\mathrm{i} \, \Delta m_{ij}^2 A_{22}'\right)\right]\\
&\quad + \sum_{i\neq j}U_{\beta i}U_{\beta j}^*U_{\alpha j}U_{\alpha i}^*\left[\cos(\Delta b_{12}^2B_{ij})-\mathrm{i} \, \sin (\Delta b_{12}^2B_{ij})\right]\exp\left(-\mathrm{i} \, \Delta m_{ij}^2A_{12}\right)\\
&\quad + \sum_{i\neq j}U_{\beta i}U_{\beta j}^*U_{\alpha j}U_{\alpha i}^*\left[\cos(\Delta b_{21}^2B_{ij})-\mathrm{i}\sin \Delta b_{21}^2B_{ij}\right]\exp\left(-\mathrm{i}\Delta m_{ij}^2A_{21}\right)\Biggr\},
\end{split}
\fe
with
\begin{align}
\label{App'}
A_{pp}' = A_{pp}-\frac{ \left( M - \frac{\Theta ^2}{64 M} \right)}{2E_0}\ln b_p^2.
\end{align}

The terms within the curly brackets in the probability expression (\ref{asdPdadbd2}) require clarification. The first term applies to cases where \(i = j\). The second term corresponds to situations with \(i \neq j\) and \(p = q\). The third and fourth terms address scenarios with \(i \neq j\) and \(p \neq q\), distinguishing between \(p < q\) and \(p > q\), respectively. For the two--flavor neutrino case, the leptonic mixing matrix is expressed as a \(2 \times 2\) matrix characterized by the mixing angle \(\alpha\) \cite{neu43}
\begin{align}
\label{U}
U\equiv\left(\begin{matrix}
\cos\alpha&\sin\alpha\\
-\sin\alpha&\cos\alpha
\end{matrix}\right).
\end{align}

Substituting the mixing matrix (\ref{U}) into Eq. (\ref{asdPdadbd2}), the oscillation probability for the transition \(\nu_{e} \to \nu_{\mu}\) is
\ie
\begin{split}
\mathcal{P}_{\alpha\beta}^{\mathrm{lens}}
&=\left|N\right|^2\sin^22\alpha\biggl\{\sin^2\dfrac{\Delta m_{21}^2A_{11}'}{2}+\sin^2\dfrac{\Delta m_{21}^2A_{22}'}{2}
+\dfrac{1}{2}\cos[\Delta b_{12}^2B_{11}]+\dfrac{1}{2}\cos[\Delta b_{12}^2B_{22} \notag\\
&\quad-\dfrac{1}{2}\cos[\Delta m_{21}^2A_{12}]
\left(\cos[\Delta b_{12}^2B_{12}] + \cos[ \Delta b_{21}^2B_{12}] \right)\notag\\
&\quad + \dfrac{1}{2} \sin [ \Delta m_{21}^2A_{12}]
\left( \sin [  \Delta b_{12}^2B_{12} ] + \sin[ \Delta b_{21}^2B_{12}]\right)\biggr\}.
\end{split}
\fe

Taking into account the leptonic mixing matrix (\ref{U}) and the phase differences associated with the different paths of neutrino propagation, the normalization constant can be expressed below:
\ie
\left|N\right|^2
=\biggl\{2+2\cos^2\alpha\cos[\Delta b_{12}^2B_{11}]
+2\sin^2\alpha\cos [\Delta b_{12}^2B_{22}]\biggl\}^{-1}.
\fe


\section{Conclusion}

This study focused on exploring various phenomena associated with a non--commutative black hole, including geodesics, matter accretion, gravitational lensing, time delay, and neutrino--related effects. Initially, the foundational aspects of the black hole under consideration were introduced, with an emphasis on encapsulating all modifications arising from $\Theta$ into the black hole's mass. Subsequently, the geodesic equations were derived numerically by solving a system of four differential equations. The outcomes were analyzed for different configurations of the non--commutative parameter \(\Theta\) and the mass \(M\), highlighting their impact on the particle trajectories.

The thin accretion disk was then investigated by considering an accretion flow, characterized as an optically thin and radiating medium consisting of infalling gas following a radial trajectory. To model the shadows produced by such a flow, a numerical framework based on the Backward Raytracing method \cite{Bambi:2012tg, Okyay:2021nnh} was employed. This approach was used to compute the shadows and raytracing, illustrating distinct photon behaviors, including direct photons, lensed photons, and the photon ring.

Gravitational lensing was analyzed through two approaches: the weak deflection angle, calculated using the Gauss–Bonnet theorem, and the strong deflection angle, determined through Tsukamoto's method. In both cases, the deflection angles, \(\alpha(b, \Theta)\) for the weak limit and \(a(b, \Theta)\) for the strong limit, increased with higher values of \(\Theta\). For the strong lensing scenario, lensing equations and observables were derived and applied to observational data of Sagittarius \( A^{*} \). Key parameters, including the critical impact parameter \(b_{c} = 3 \sqrt{3} \left(M - \frac{\Theta^2}{64 M} \right)\), and the angular size, \(\theta_{\infty} \approx 25.24 \, \mu\text{arcsecs} + \mathcal{O}(\Theta^{2})\), were computed.

Also, the study addressed the time delay, energy deposition rate from neutrino pair annihilation, neutrino oscillation phases and probabilities, and the effects of gravitational lensing on neutrinos. Additionally, exploring neutrino--related phenomena under varying implementations of non--commutativity, such as Lorentzian or Gaussian distributions, presents an interesting direction for future research.



\section*{Acknowledgments}
\hspace{0.5cm} A. A. Araújo Filho is supported by Conselho Nacional de Desenvolvimento Cient\'{\i}fico e Tecnol\'{o}gico (CNPq) and Fundação de Apoio à Pesquisa do Estado da Paraíba (FAPESQ), project No. 150891/2023-7. A.{\"O} and N.H. would like to acknowledge the network support by the COST Action CA21106 - COSMIC WISPers in the Dark Universe: Theory, astrophysics and experiments (CosmicWISPers), the COST Action CA21136 - Addressing observational tensions in cosmology with systematics and fundamental physics (CosmoVerse), the COST Action CA22113 - Fundamental challenges in theoretical physics (THEORY-CHALLENGES) and the COST Action CA23130 - Bridging high and low energies in search of quantum gravity (BridgeQG) and the COST Action CA23115 - Relativistic Quantum Information (RQI) funded by COST (European Cooperation in Science and
Technology). We also thank EMU, TUBITAK ULAKBIM, Turkiye and SCOAP3, Switzerland for their support.

\section{Data Availability Statement}

Data Availability Statement: No Data associated in the manuscript

\bibliographystyle{ieeetr}
\bibliography{main}

\end{document}